%% file: Manuscript.tex
\documentclass[aps,prl,reprint,superscriptaddress,nobibnotes]{revtex4-1}

\usepackage{graphicx}
\usepackage{setspace}
\usepackage{esint}
\usepackage{multibib}
   \newcites{S}{Supplementary}
\usepackage{array}

\begin{document}

\title{Edge-mode Superconductivity in a Two Dimensional Topological Insulator}



\author{Vlad S. Pribiag}
	\altaffiliation{These authors contributed equally to this work.}
\author{Arjan J.A. Beukman}
	\altaffiliation{These authors contributed equally to this work.}
\author{Fanming Qu}
	\altaffiliation{These authors contributed equally to this work.}
\author{Maja C. Cassidy}
\affiliation{Kavli Institute of Nanoscience, Delft University of Technology, 2600 GA Delft, The Netherlands}
\author{Christophe Charpentier}
\author{Werner Wegscheider}
\affiliation{Solid State Physics Laboratory, ETH Z\"urich, 8093 Z\"urich, Switzerland}
\author{Leo P. Kouwenhoven}
	\email{l.p.kouwenhoven@tudelft.nl}
\affiliation{Kavli Institute of Nanoscience, Delft University of Technology, 2600 GA Delft, The Netherlands}
	




\date{\today}

\begin{abstract} 
Topological superconductivity is an exotic state of matter that supports Majorana zero-modes, which are surface modes in 3D, edge modes in 2D or localized end states in 1D \cite{01_Hasan2010,02_Qi2011}. 
In the case of complete localization these Majorana modes obey non-Abelian exchange statistics making them interesting building blocks for topological quantum computing \cite{03_Nayak2008,04_Read2012}. 
Here we report superconductivity induced into the edge modes of semiconducting InAs/GaSb quantum wells, a two-dimensional topological insulator \cite{05_Kane2005,06_Bernevig2006,07_Bernevig2006a,08_Konig2007,09_Liu2008,10_Du2013}. 
Using superconducting quantum interference, we demonstrate gate-tuning between edge-dominated and bulk-dominated regimes of superconducting transport.
The edge-dominated regime arises only under conditions of high-bulk resistivity, which we associate with the 2D topological phase. 
These experiments establish InAs/GaSb as a robust platform for further confinement of Majoranas into localized states enabling future investigations of non-Abelian statistics.
\end{abstract}


\maketitle

Several studies have reported topological superconductivity in 3D \cite{11_Sasaki2011} and 1D \cite{12_Mourik2012,13_Das2012,14_Deng2012,15_Churchill2013} materials. 
In 2D semiconductor quantum wells a topological insulator (TI) is identified by the observation of a quantum spin Hall effect \cite{05_Kane2005,06_Bernevig2006}. 
In this phase the 2D bulk is a gapped insulator and transport only occurs in gapless edge states. 
These edge modes are counter-propagating, spin-polarized channels, known as helical modes, which are protected against elastic backscattering in the presence of time-reversal symmetry. 
To date, only two 2D TI systems have been identified experimentally: HgTe/HgCdTe quantum wells \cite{08_Konig2007} and InAs/GaSb double quantum wells \cite{10_Du2013,11_Sasaki2011,12_Mourik2012,13_Das2012,14_Deng2012,15_Churchill2013,16_Knez2011a}. 
The origin of the TI phase is different for the two materials: relativistic band-bending for HgTe/HgCdTe \cite{07_Bernevig2006a} and type-II broken band alignment for InAs/GaSb \cite{09_Liu2008}. 
Recent scanning microscopy experiments have confirmed the presence of edge currents in both 2D TIs \cite{17_Nowack2013,18_Spanton2014}. 
The two different material classes are considered interesting complementary alternatives for topological studies.

\begin{figure}
	\includegraphics{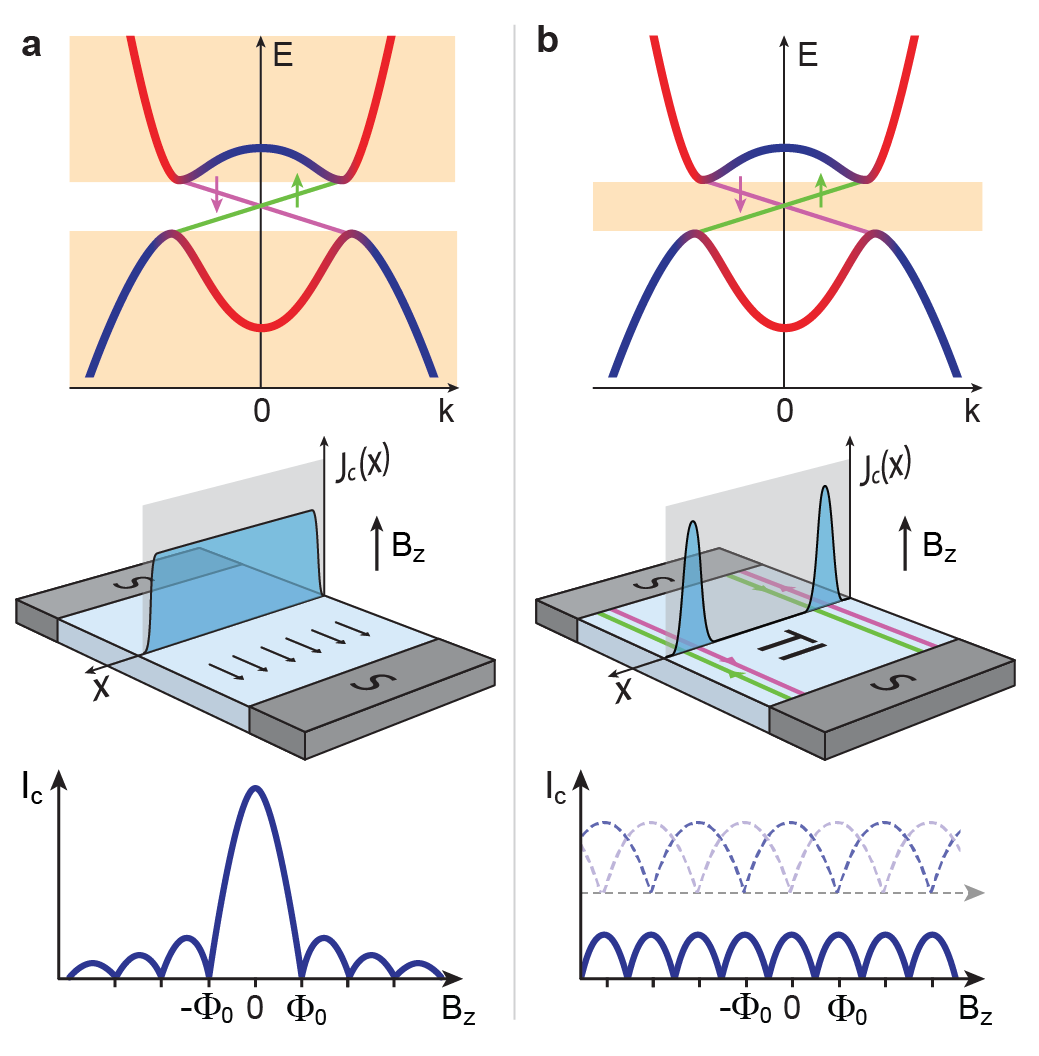}
	\caption{\label{fig1} Band structure and SQI patterns. The top panel shows schematic band diagrams for InAs/GaSb quantum wells. Due to the type-II broken band alignment within the heterostructure, the electron (red) and hole (blue) 2D-bulk bands cross. Coupling between these bands opens up a topological gap, which is crossed by gapless, linearly-dispersive helical edge states. (a), When the Fermi level is in one of the bulk bands (coloured rectangles) the critical current density profile is spatially uniform (middle panel) and the corresponding SQI has a Fraunhofer-like shape with a central lobe of width $2\Phi_0$ and side lobes of width $\Phi_0$ (bottom panel). (b), When the Fermi level is in the topological gap and crosses the helical edge modes (coloured rectangle), the current density profile is localized at the edges (middle panel) and the corresponding SQI has a SQUID-like shape (bottom panel). A $2\Phi_0$-periodic SQI is expected for the helical edge modes in the absence of quasiparticle poisoning (2 phases are possible, as shown by the dashed lines in the bottom panel, depending on whether or not the two edges have the same fermion parity). Quasiparticle poisoning can induce fermion parity switches that restore the $\Phi_0$ periodicity even for helical modes (bottom panel, solid line).}
\end{figure}

Effects from proximitizing 2D TIs with superconductors have been investigated, such as excess currents due to Andreev reflection \cite{19_Knez2012} and Josephson effects in SNS (superconductor-normal-superconductor) junctions \cite{20_Oostinga2013}, as illustrated in Fig. 1(a).
To demonstrate topological superconductivity (TS), however, it needs to be shown that superconducting transport takes place along the helical edges, as depicted in Fig. 1(b). 
Here we demonstrate explicitly edge-mode superconductivity in InAs/GaSb. 
Recently, a similar experiment was reported by Hart \textit{et al.} in the HgTe material \cite{21_Hart2014}. Below, we discuss the topological and helical aspects of this edge-mode superconductivity.

A straightforward consequence of the conventional SNS junction in contrast to an edge-mode superconducting junction can be observed in a superconducting quantum interference (SQI) measurement, where a perpendicular magnetic field induces oscillations in the amplitude of the superconducting current. 
A conventional SNS junction yields the Fraunhofer pattern, as shown in the bottom panel of Fig. 1(a). 
In the case of edge-mode superconductivity only the junction effectively acts as a SQUID (superconducting quantum interference device) with a well-known $\Phi_0$-periodic interference pattern, see bottom panel of Fig. 1(b).

To specify this further, we consider a short Josephson junction (defined as $L \ll \zeta$, where $L$ is the contact separation and $\zeta = \hbar v / \Delta_{\text{ind}}$ is the superconducting coherence length in the junction material with Fermi velocity $v$ and induced gap $\Delta_{\text{ind}}$) which has a sinusoidal current-phase relation. 
In this case, the Josephson supercurrent, $I_{\text{s}}(B_{\text{z}})$, is given by the Fourier transform of the density profile of the critical current, $J_{\text{c}}(x)$, taken at a perpendicular magnetic field $B_{\text{z}} = 0$:
$$
I_{\text{s}}(B_{\text{z}}) = \text{Im} \left[ \int_{-\infty}^{ \infty} J_{\text{c}} (x) e^{ikx + \phi_0} dx \right],
$$ 
with the effect of magnetic field included in $k = 2 \pi L B_{\text{z}} / \Phi_0$ \cite{22_Dynes1971}, and where $\phi_0$ is the superconducting phase difference between the contacts.
The critical current becomes 
$$
I_c(B_{\text{z}}) \equiv \max \left[ I_{\text{s}} (B_{\text{z}}) \right] = \left\Vert \int_{-\infty}^{\infty} J_{\text{c}}(x) e^{ikx} dx \right\Vert.
$$
For a spatially-uniform $J_{\text{c}}(x)$ = constant, the SQI pattern has the typical Fraunhofer form, $\left| \sin(\pi L W B_{\text{z}}/\Phi_0)/(\pi L W B_{\text{z}}/\Phi_0) \right|$, with a central lobe of width $2\Phi_0$ and side lobes of width $\Phi_0$ ($\Phi_0 = h/2e$ is the superconducting flux quantum), see Fig. 1(a).
In contrast, for edge-mode superconductivity, the SQI is simply $\Phi_0$-periodic (see Fig. 1(b)). 
Note that this analysis does not include effects with topological origin, such as when the edge modes have helical character. 
In that case the SQI can become $2\Phi_0$-periodic \cite{01_Hasan2010,02_Qi2011}, as illustrated in Fig. 1(b) and discussed later in the paper.

\begin{figure}
	\includegraphics{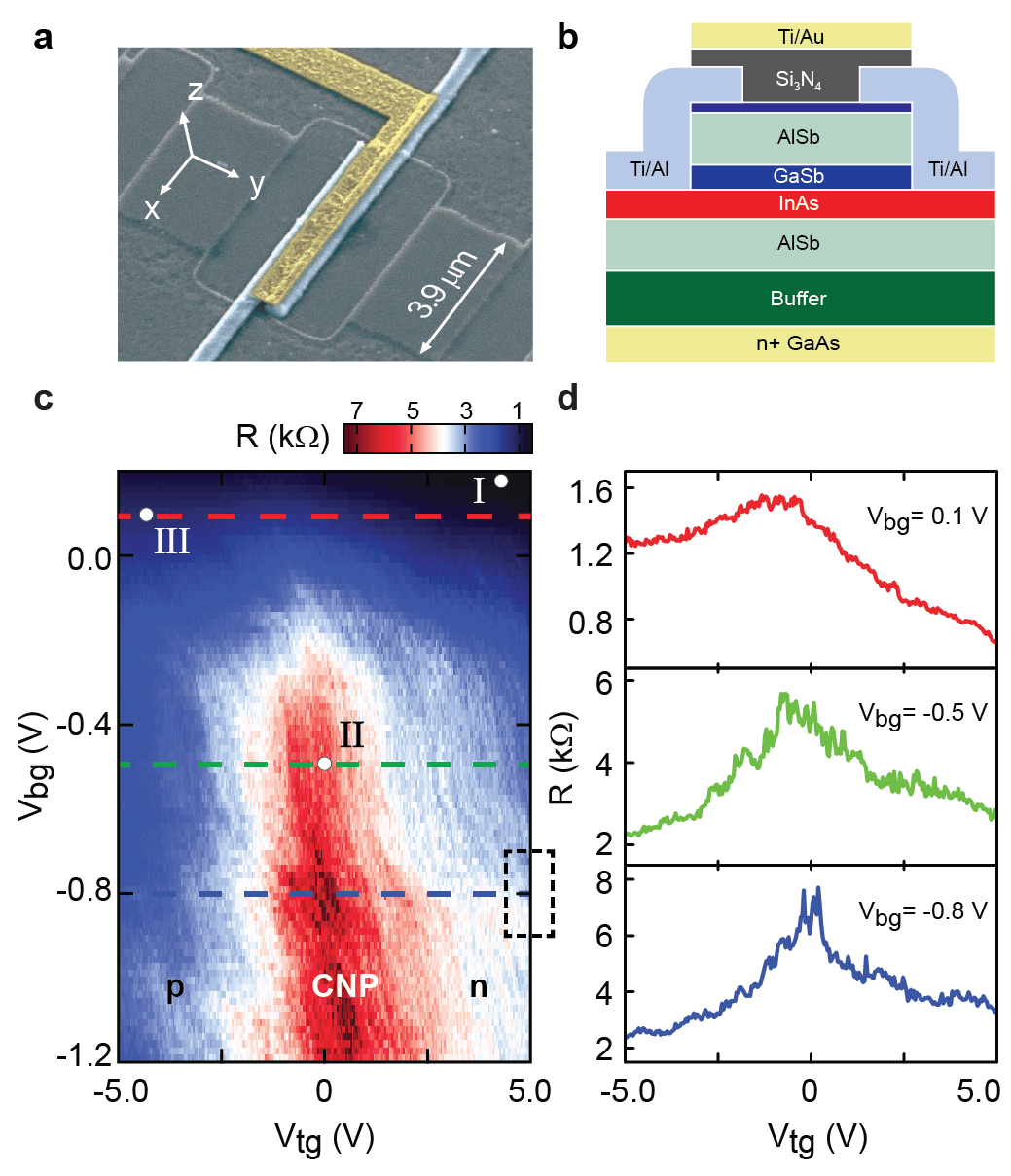}
	\caption{\label{fig2} Device layout and normal state transport. (a), False colour scanning electron microscope image of a typical S-InAs/GaSb-S junction. The superconducting material, S, is Ti(5 nm)/Al(150 nm) (see Supplementary Figure S12 for devices with $\text{NbTiN}_{\text{x}}$ contacts). (b), Cross-sectional view of device layout. (c), Phase diagram measured on InAs/GaSb (device A, cooldown 1). $R_{\text{N}}$ is measured using a DC excitation current $I_{\text{sd}}$ = 5 nA. The Ti/Al contacts are driven into the normal state by an applied field $B_{\text{z}}$ = 100 mT. The dashed rectangle refers to the data discussed in Fig. 5. (d), Line cuts showing $R_{\text{N}}$ as a function of $V_{\text{tg}}$ for three different $V_{\text{bg}}$ values (corresponding to the dashed lines in c).}
\end{figure}

\begin{figure*}
	\includegraphics{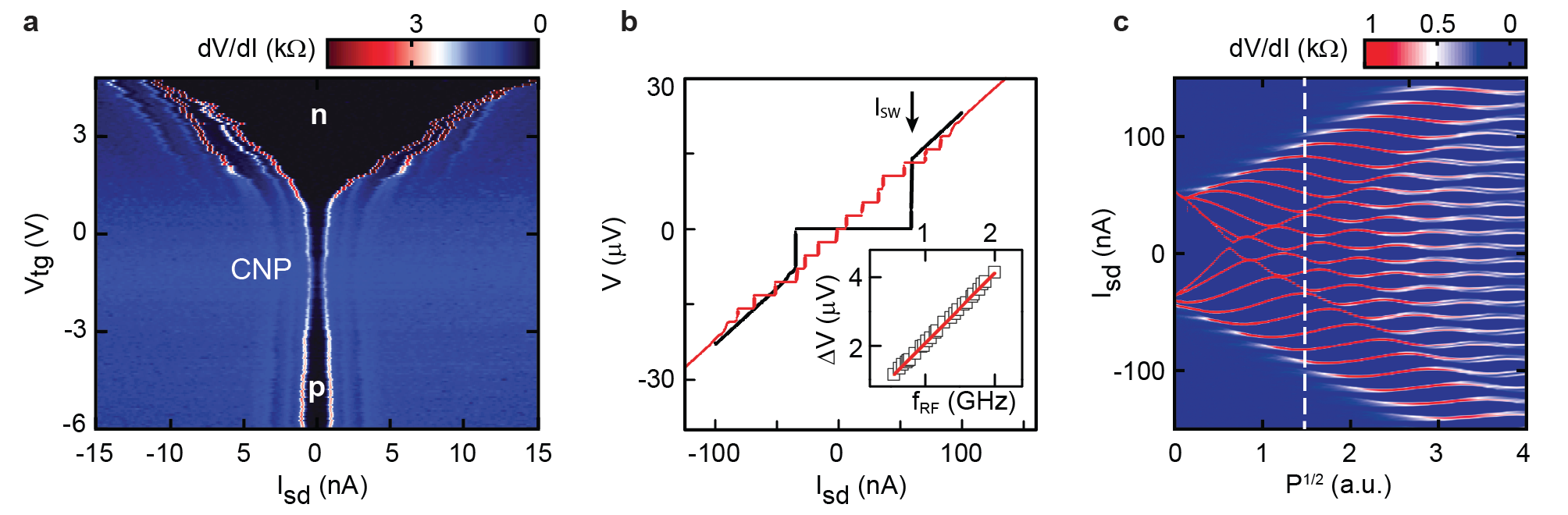}
	\caption{\label{fig3} Josephson effect (device A, cooldown 2). (a), $dV/dI$ vs. $I_{\text{sd}}$ and $V_{\text{tg}}$ at $B=0$, showing gate-tunable supercurrent through the junction ($V_{\text{bg}}$ = 0.1 V). The three main transport regions are indicated by the labels n (Fermi level in the conduction band), p (Fermi level in the valence band) and CNP (Fermi level at the charge neutrality point). (b), I-V traces without microwaves (black) and with microwaves (red), with $f_{\text{RF}}$= 1.288 GHz. Inset: frequency-dependence of the Shapiro step height, showing the expected linear dependence. (c), Dependence of the Shapiro plateaus on microwave field amplitude, $P^{1/2}$, for $V_{\text{tg}}$ = 5 V and $V_{\text{bg}}$ = 0.2 V. The white dashed line indicates the line cut corresponding to the red curve in (b).}
\end{figure*}

Before investigating the superconducting regime, we first describe normal state transport in our Ti/Al-InAs/GaSb-Ti/Al junctions (details of the device geometry are shown in Fig. 2(a) and 2(b)). 
We focus on one device (device A) and map out the normal state resistance, $R_{\text{N}}$, when superconductivity is suppressed by a $B_{\text{z}}$ = 0.1 T (Fig. 2(a)). 
The junction has width $W$ = 3.9 $\mu$m and contact separation $L$ = 400 nm, significantly shorter than the edge mode decoherence length of ~2 to 4 $\mu$m \cite{10_Du2013,16_Knez2011a}. 
Transport is gate-tuned using the $\textrm{n}^+$ GaAs substrate as a back gate, and a Ti/Au top gate. 
As the topgate voltage, $V_{\text{tg}}$, is tuned from positive to negative, a resistance peak develops indicating a charge neutrality point (CNP) \cite{16_Knez2011a,23_Nichele2014} when the Fermi energy is located in the topological gap (see upper panel in Fig. 1(b)). 
For more positive $V_{\text{tg}}$, the Fermi level is moved up into the conduction band and the dominant charge carriers are electrons, while for more negative $V_{\text{tg}}$ the Fermi level is moved down into the valence band and charge transport is dominated by holes. 
This interpretation is confirmed by measurements in the quantum Hall regime performed on material from the same growth batch \cite{23_Nichele2014}. 
The position of the CNP shifts to more positive $V_{\text{tg}}$ as the back gate voltage, $V_{\text{bg}}$, is tuned more negative, as shown in the line cuts in Fig. 2(b), in qualitative agreement with band structure calculations \cite{09_Liu2008}. 
The maximum resistance at the CNP is $\sim 7 \ \textrm{k}\Omega$. This value is smaller than the ideal quantized value of $h/2e^2$ ($\sim 13 \ \textrm{k}\Omega$) expected for transport only via helical edge modes, indicating some residual bulk conductivity.

For $B_{\text{z}} <$ 11 mT we observe a supercurrent, a direct consequence of the DC Josephson effect. We define the switching current, $I_{\text{SW}}$, as the value of the applied bias current when the developed voltage jumps from virtually zero to a finite value (see Fig. 3(b)).
$I_{\text{SW}}$ is tuned by means of gate voltages: as $V_{\text{tg}}$ becomes less positive $I_{\text{SW}}$ first decreases, then saturates at a minimum value for $V_{\text{tg}}$ near the CNP, and then increases again for more negative $V_{\text{tg}}$ due to hole-mediated transport through the bulk (Fig. 3(a)).
To unambiguously establish the Josephson nature of our junctions, we irradiate the device with microwaves of frequency $f_{\text{RF}}$. 
We observe the familiar Shapiro ladder \cite{24_Thinkham1996} with steps at $V = nhf_{\text{RF}}/2e$ ($n = 1,2, \dots$). Fig. 3(b) shows a particular comparison of I-V curves measured without and with the microwaves, the latter showing the characteristic Shapiro steps, which are a consequence of the AC Josephson effect. 
The step heights exhibit the expected linear dependence when $f_{\text{RF}}$ is varied (inset of Fig. 3(b)). Fig. 3(c) shows the characteristic modulation of the widths of the Shapiro steps by the magnitude of the applied microwave field. Similar data near the CNP is shown in Supplementary Figure S8.

Having established the DC and AC Josephson effect in our InAs/GaSb junctions, we next analyze the spatial distribution of the supercurrent by performing SQI measurements at different gate values, see Fig. 4. 
As shown by Dynes and Fulton \cite{22_Dynes1971}, the current density profile $J_{\text{c}}(x)$ can be determined from the measured SQI provided the phase of the complex Fourier transform can be reliably estimated. This method was recently used to establish induced superconductivity in the edge modes of HgTe/HgCdTe \cite{21_Hart2014}. 
We first comment on the validity of the Dynes and Fulton approach for our devices. The superconducting coherence length for an edge mode velocity $v \approx 4.6 \cdot 10^4$ m/s in InAs/GaSb \cite{25_Wang2013} is $\zeta \geq 240$ nm (using $\Delta_{\text{ind}} \leq \Delta \approx 125 \ \mu$eV, with $\Delta$ the superconducting gap of the electrodes, see Supplementary Figure S9). 
We have verified that in our limit (where $L$ is of order $\zeta$) the SQI pattern is only weakly sensitive to deviations from a perfect sinusoidal I-$\Phi$ relation, so the Dynes and Fulton short junction approach is indeed justified. 

\begin{figure*}[!]
	\includegraphics{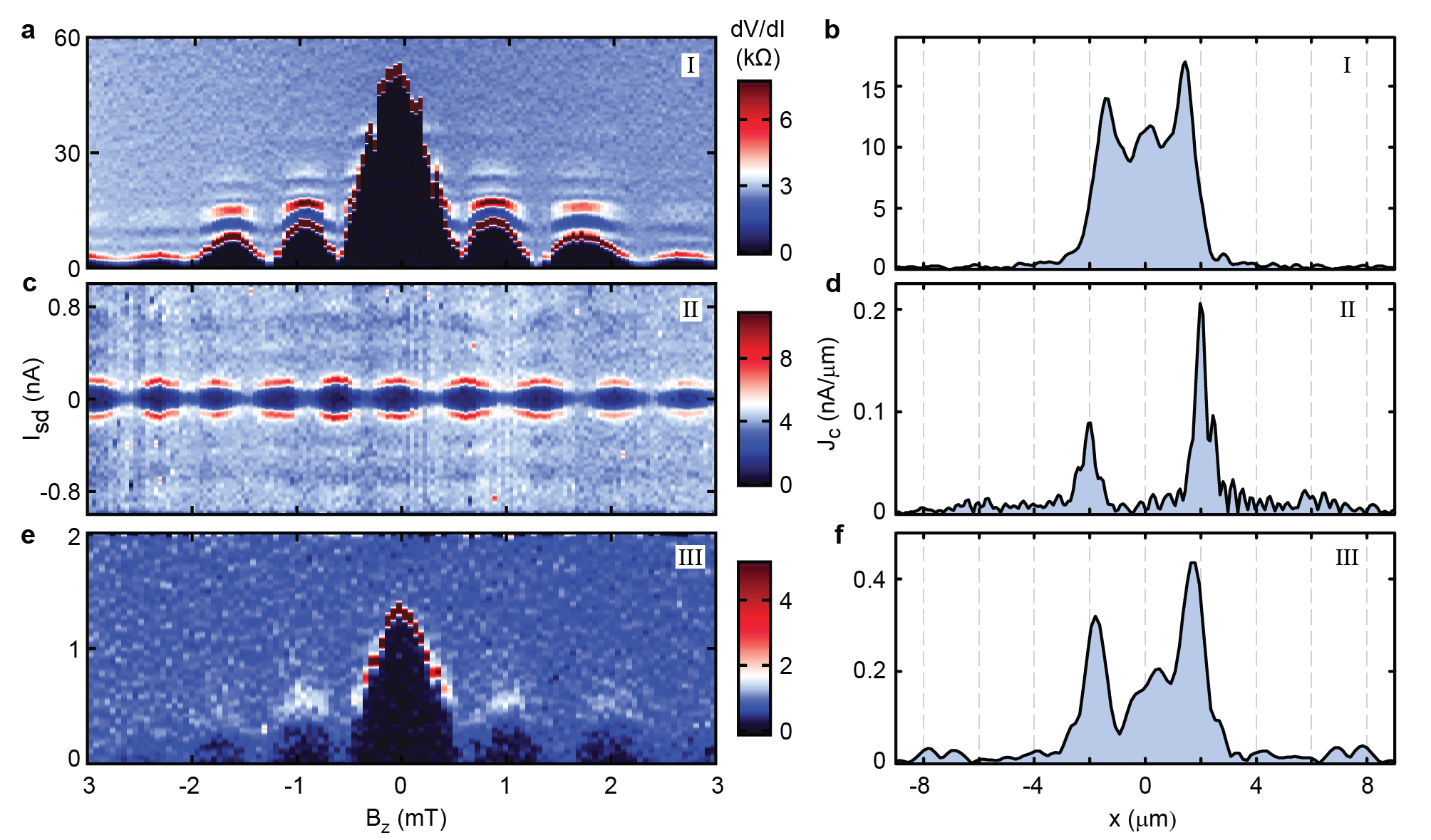}
	\caption{\label{fig4} Superconducting quantum interference (SQI) patterns and corresponding current density profiles. (a),(b), In the n-region ($V_{\text{tg}} = 4.8$ V and $V_{\text{bg}} = 0.2$ V). (c),(d), At the CNP ($V_{\text{tg}} = -0.3$ V and $V_{\text{bg}} = -0.4$ V). (e),(f), In the p-region ($V_{\text{tg}} = -4.8$ V and $V_{\text{bg}} = 0.15$ V). Data was measured on device A (cooldown 1). The gate values are indicated (I-III) in Fig. 2(c). The effective device area used to extract $J_{\text{c}}(x)$ was determined by requiring that the nodes of the SQI pattern be at multiples of $\Phi_0$. Given the lithographic width, $W = 3.9 \ \mu$m, we compute an effective junction length $L_{\text{eff}} \sim 640$ nm. This is longer than the contact separation, $L = 400$ nm, due to flux focusing by the superconducting contacts.}
\end{figure*}

Figure 4 summarizes our main result: gate-tuning from bulk to edge-mode superconductivity. The figure shows SQI data at representative points in gate space indicated in Fig. 2(c), along with the current density profiles extracted using the Dynes and Fulton approach \cite{21_Hart2014,22_Dynes1971}.
We observe three regimes: I) a distinct Fraunhofer pattern when the Fermi energy is in the conduction band. 
The corresponding current density profile indicates that most of the current is carried by the bulk (Fig. 4(a),(b)). 
II) a SQUID-like interference when the Fermi energy is near the CNP. In this regime, the supercurrent density is clearly edge-mode dominated (Fig. 4(c),(d)).
III) A return to a Fraunhofer-like pattern as the Fermi energy enters the valence band. Here, the current distribution acquires a large bulk contribution, but edge modes also contribute over the range of accessible gate voltage values (Fig. 4(e),(f)).
Supplementary Figures S3-5 include additional SQI patterns measured at other points within gate space. 
Taken together, these data clearly demonstrate gate tuning between bulk and edge-mode superconductivity in InAs/GaSb. 
As a further check, we studied a non-topological InAs-only junction (device B), where, as expected, a SQUID-like SQI was not observed (see Supplementary Figure S11).

The edge-mode SQI data typically shows conventional $\Phi_0$-periodicity, e.g. as in Fig. 4(c). 
However, over a certain gate range (see dashed rectangle in Fig. 2(c)) we observe a striking even-odd pattern in the interference lobes. 
An example is shown in Fig. 5(a). This $2\Phi_0$-periodic effect is also seen in another device with different contact material (see Supplementary Figure S12). 
To the best of our knowledge, this observation can have two types of interpretations, one conventional and one that explicitly requires including topological effects. 
First, the conventional Dynes and Fulton analysis would require a current density profile containing three peaks, two at the edges and one in the middle (see Fig. 5(b)).
Simulations of such $2\Phi_0$-SQI (Supplementary Figure S13) indicate that this conventional analysis would require a third channel that is within 10\% of the device center. 
It is improbable that such an effect would occur in two separate devices. 
A second, alternative, explanation is that the $2\Phi_0$-periodicity could instead originate from what is known as the fractional Josephson effect \cite{26_Fu2009}. 
In this interpretation, the edge modes contain Majorana zero-modes. Josephson-coupled Majoranas transport a charge $e$, instead of the conventional $2e$ Cooper pair charge, resulting in a doubling of the SQI periodicity \cite{27_Beenakker2013,28_Lee2014}. 
This interpretation, however, requires a quasiparticle poisoning time scale that is in excess of the measurement time (tens of seconds). Using existing techniques \cite{29_Nguyen2013}, future experiments should directly measure quasiparticle poisoning to further establish the topological nature of this $2\Phi_0$-periodic effect.

\begin{figure*}
   	\includegraphics{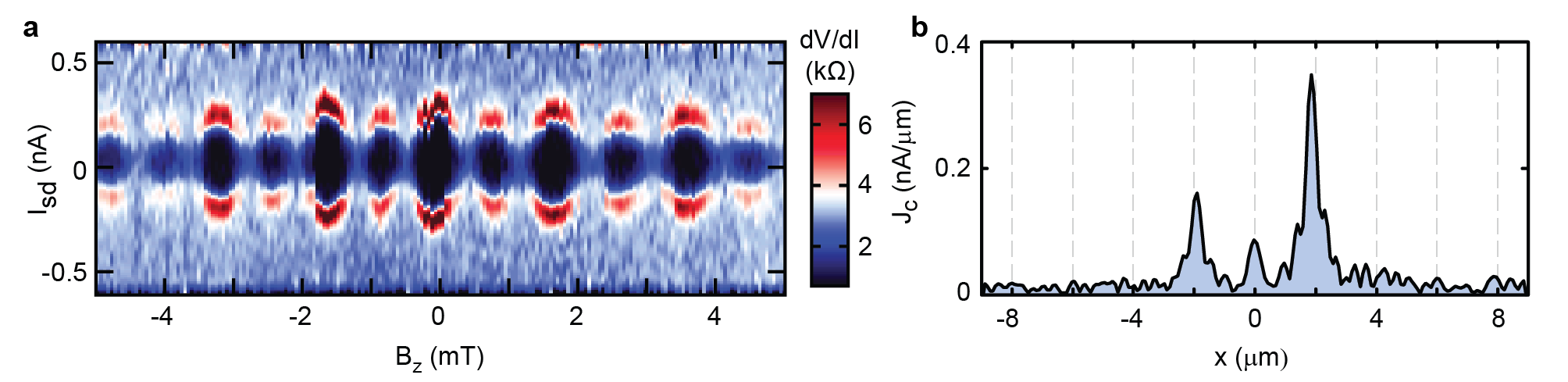}
   	\caption{\label{fig5} $2\Phi_0$-periodic quantum interference pattern. (a), SQI pattern measured at $V_{\text{tg}} = 5.5$ V and $V_{\text{bg}} = -0.8$ V, showing a pronounced even-odd effect with $2\Phi_0$ periodicity. (b), The corresponding current density profile assuming a conventional I-$\Phi$ relation.}
\end{figure*}

\textbf{Methods: } 
The InAs/GaSb quantum wells were grown using Molecular Beam Epitaxy on $\textrm{n}^+$ (001) GaAs substrates. 
Two different material batches were used: a batch grown using high mobility Ga (HM) and a batch using lower-mobility Ga (LM). 
The LM batch has lower residual bulk conductance near the CNP. Measurements were performed in a dilution refrigerator with a mixing chamber temperature of 16 mK equipped with a three-axis vector magnet. SQI patterns corresponding to an edge-mode current density profile were observed in three devices: device A from the main text (HM heterostructures and Al contacts), and devices C and D (based on LM heterostructures and with $\text{NbTiN}_{\text{x}}$ contacts, see Supplementary Figure S12). Device A was measured in two separate cooldowns. No significant changes in the device properties were observed between cooldowns. Offsets in $B_{\text{z}}$ of up to a few mT due to trapped flux in the superconducting magnets or leads were subtracted in the plotted SQI data. The spatial resolution of the current density profiles extracted from SQI patterns is $\sim W\Phi_0 / \Delta \Phi$, where $\Delta \Phi$ is the magnetic flux range of the SQI measurement. In each of the plots, the FWHM of the InAs/GaSb edge modes is near the Fourier resolution limit and represents an upper bound on the actual width of the edge mode. The maximum $\Delta \Phi$ is limited by reduced visibility of the oscillations for $B_{\text{z}} \geq$ 11 mT in the case of Al contacts, and by switches along the $B_{\text{z}}$ axis in the case of $\text{NbTiN}_{\text{x}}$ contacts (presumably due to flux depinning in the leads, see Supplementary Figure S12).

\begin{acknowledgments}
The authors thank A. Akhmerov, D. Pikulin, M. Wimmer, T. Hyart, C. Beenakker and A. Geresdi for valuable discussions and comments, and K. Zuo for assistance with the dilution refrigerator. This work has been supported by funding from the Netherlands Foundation for Fundamental Research on Matter (FOM) and Microsoft Corporation Station Q.  V.S.P. acknowledges funding from the Netherlands Organisation for Scientific Research (NWO) through a VENI grant. Two of the authors (C.C. and W.W.) acknowledge funding by the Swiss National Science Foundation (SNF). 
\end{acknowledgments}

V.S.P., A.J.A.B. and F.Q. fabricated the devices and performed the measurements. C.C. and W.W. provided the InAs/GaSb heterostructures. V.S.P., A.J.A.B., F.Q., M.C.C. and L.P.K. contributed to the experiments and all authors discussed the results and edited the manuscript.

\bibliography{Manuscript}

\include{Supplementary}

\end{document}

%% file: Supplementary.tex
\onecolumngrid
\setcounter{figure}{0}

\makeatletter 
\renewcommand{\thefigure}{\normalsize\textbf{S\arabic{figure}}}
\renewcommand{\thetable}{\normalsize \textbf{S\arabic{table}}}
\makeatother

\renewcommand\figurename{\normalsize \textbf{Figure}}
\renewcommand\tablename{\normalsize\textbf{Table}}

\newpage

\centering
\LARGE{ \huge Supplementary Information \\[5mm] \LARGE Edge-mode Superconductivity in a Two Dimensional Topological Insulator}\\[2cm]

\large List of Supplementary Figures\\[2cm]

\large
\begin{tabular}{p{2.5cm} p{13cm}}
Figure S1: 	& Device images and measurement setup. \\[5mm]
Table S2:	& Details of the devices discussed in the main text and supplementary information.\\[5mm]	
Figure S3:	& Gate-dependence of SQI patterns.\\[5mm]
Figure S4:	& Even-odd effect in the switching current.\\[5mm]
Figure S5:	& SQI pattern over a large magnetic field range.\\[5mm]
Figure S6:	& Magnetic-field dependence of the normal-state resistance.\\[5mm]
Figure S7:	& Temperature-dependence of the critical current.\\[5mm]
Figure S8:	& Shapiro steps near the charge neutrality point.\\[5mm]
Figure S9:	& Line-cuts showing representative I-V curves in the three transport regimes.\\[5mm]
Figure S10:	& Gate-dependence of the switching current.\\[5mm]
Figure S11:	& Superconducting and normal transport for an S-InAs-S junction.\\[5mm]
Figure S12:	& SQI patterns for an S-InAs/GaSb-S junction based on InAs/GaSb grown using a lower mobility Ga source and contacted with $\text{NbTiN}_{\text{x}}$.\\[5mm]
Figure S13:	& Simulated SQI patterns.
\end{tabular}

\newpage
\linespread{1.5}

\begin{figure}
   	\includegraphics[width=150mm]{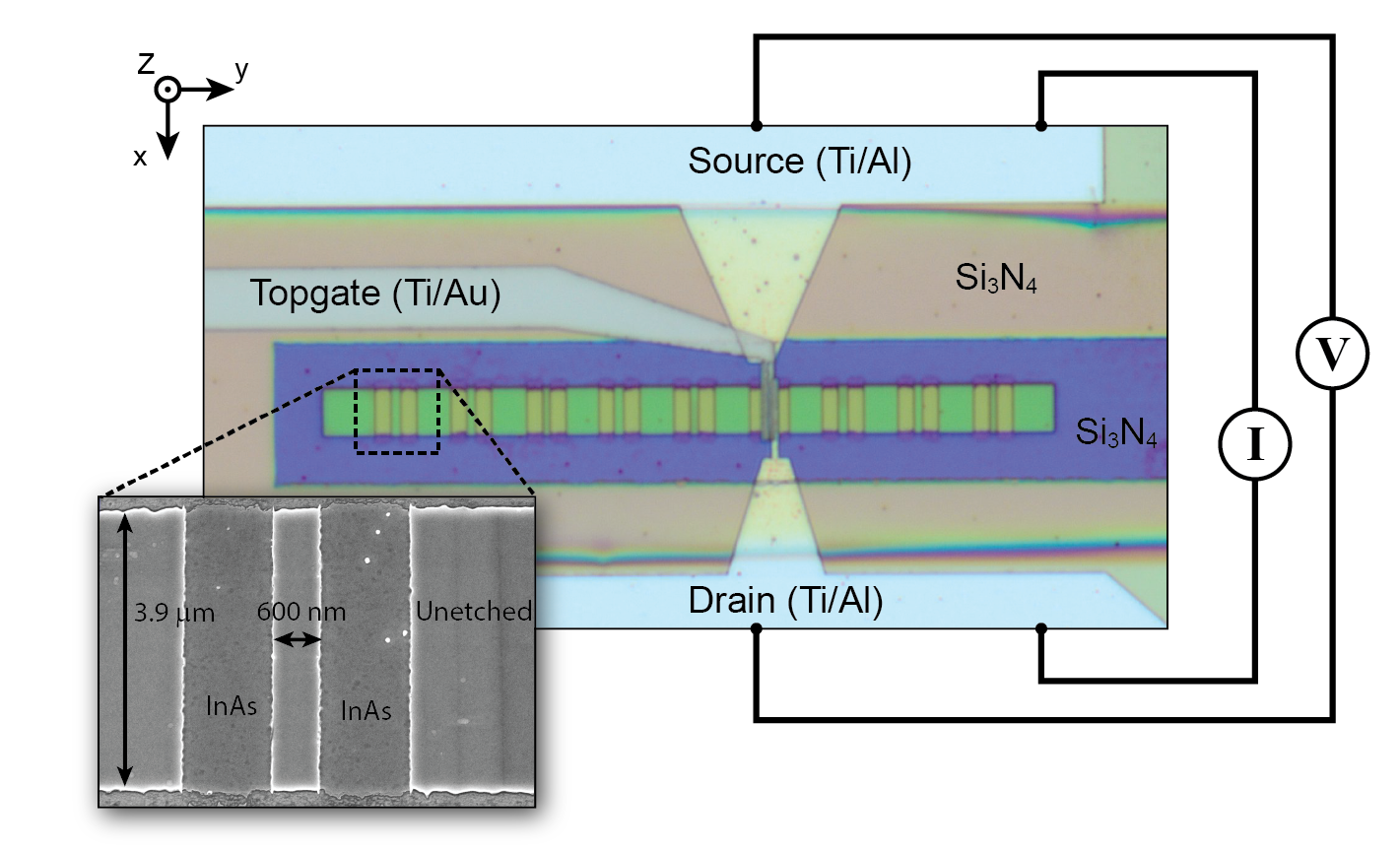}
   	\caption{\label{sup1} \linespread{1.5} \normalsize \textbf{Device images and measurement setup}. Optical microscope image of a completed S-InAs/GaSb-S junction, similar to device A discussed in the main text. The inset shows a scanning electron microscope image for the InAs/GaSb mesa, defined using electron beam lithography and wet etching. The mesa was isolated by a selective wet etch that stops at the 50 nm-thick AlSb barrier (see heterostructure schematic in Fig. 2(b) in the main text). The stack was also selectively wet-etched down to the InAs layer, leaving behind an unetched ridge (600 nm-wide in the inset). Ridges were selected based on optical inspection and contacted by depositing Ti (5 nm)/Al (150 nm) onto the exposed InAs layer using e-beam evaporation. This was followed by sputtering a 100 nm-thick $\textrm{Si}_3\textrm{N}_4$ gate dielectric layer and evaporating the Ti/Au top gate. For the measurements, two wires were bonded to each Ti/Al contact, enabling quasi-four-terminal measurements with separate current and voltage wires from room temperature down to the device bonding pads.}
\end{figure}

\setcounter{table}{1}
\begin{table}
\begin{tabular}{|l|>{\centering}p{18mm}|>{\centering}p{32mm}|>{\centering}p{25mm}|c|}
\hline
Device name & Width (W) ($\mu$m) & Contact spacing (L) (nm) & Contact material & Heterostructure type \\ \hline
A			& 3.9		& 400 & Ti/Al & HM Ga \\ \hline
B (InAs)	& 3.9		& 400 & Ti/Al & HM Ga \\ \hline
C			& 3.9		& 450 & $\textrm{NbTiN}_{\text{x}}$ & LM Ga \\ \hline
D			& 3.9		& 550 & $\textrm{NbTiN}_{\text{x}}$ & LM Ga \\ \hline
\end{tabular}
\caption{\large \textbf{Details of the devices discussed in the main text and Supplementary Information}}
\end{table}

\setcounter{figure}{2}

\begin{figure}
   	\includegraphics[width=\linewidth]{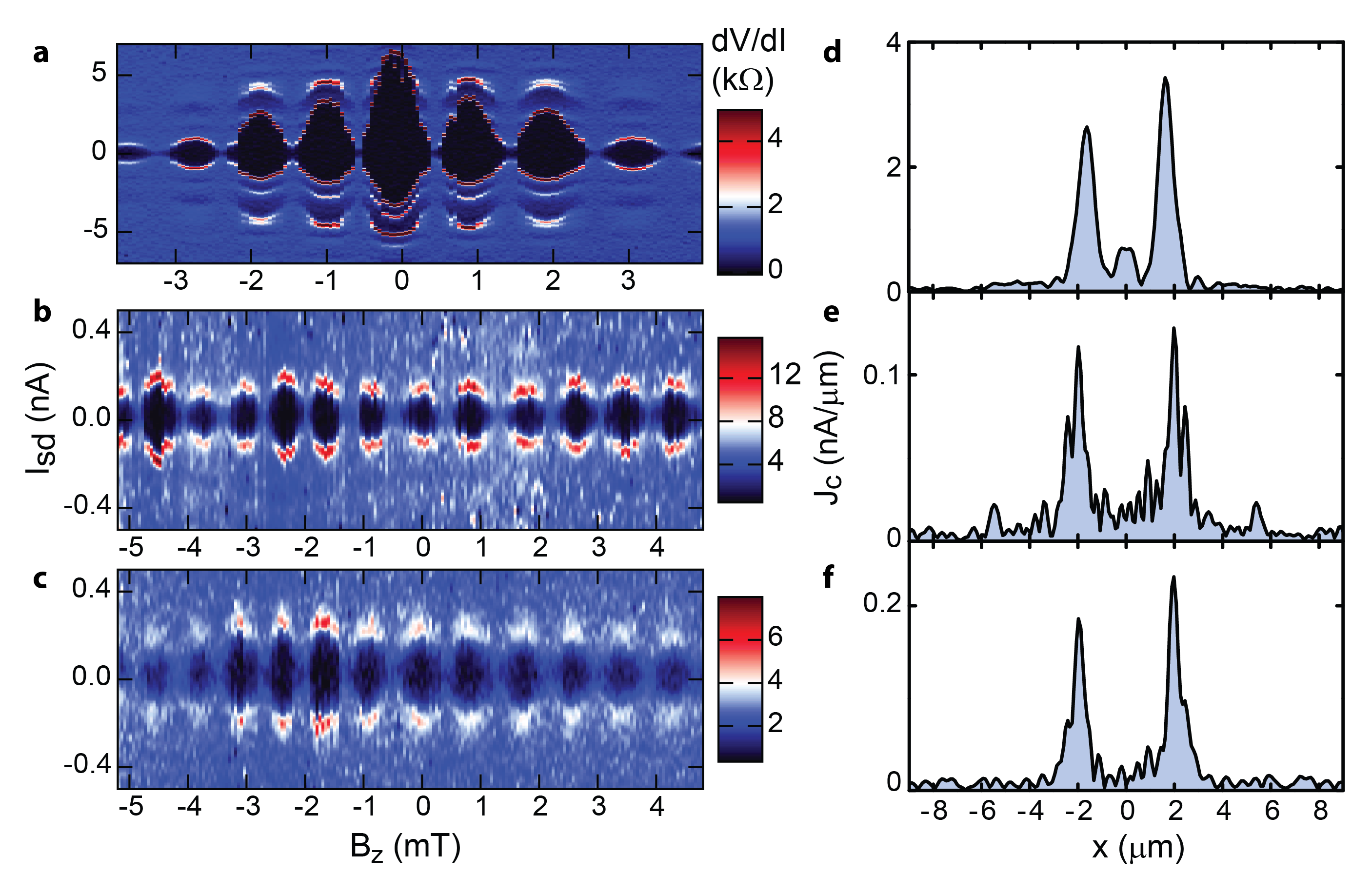}
   	\caption{\label{sup3} \linespread{1.5} \normalsize \textbf{Gate-dependence of SQI patterns.} (a)-(c), Differential resistance, $dV/dI_{\text{sd}}$, as a function of perpendicular magnetic field, $B_{\text{z}}$, and source-drain current bias, $I_{\text{sd}}$, for several bottom and top gates settings (device A).
   	(a), For ($V_{\text{bg}}$, $V_{\text{tg}}$) = (0, 4.8) $V$ the device has a normal state resistance $R_{\text{N}} \sim 900 \ \Omega$, an intermediate value between the deep electron regime and the charge neutrality point. The corresponding current density profile obtained by reverse Fourier transform \cite{01_Dynes1971,02_Hart2014} (as described in the main text) is shown in (d). It indicates supercurrent contributions from both the bulk (c.f. Fig. 4(a) in the main text) and the edge modes (c.f. Fig. 4(b) in the main text). 
   	(b), For $(V_{\text{bg}}, V_{\text{tg}}) = (0, -0.8)$ V the device is in the CNP regime, with $R_{\text{N}} \sim 5000 \ \Omega $. The data shows a SQUID-like SQI pattern. The corresponding current density profile in (e) is dominated by the edge modes. 
   	(c), For $(V_{\text{bg}}, V_{\text{tg}}) = (-0.8, -4.8)$ V the device is in the hole regime, with $R_{\text{N}} \sim 2600 \ \Omega $. 
   	The corresponding current density profile is shown in (f). In general, we observe that the SQI pattern becomes SQUID-like, corresponding to edge-mode-dominated superconducting transport, whenever $R_{\text{N}} \geq 900 \ \Omega$.}
\end{figure}

\begin{figure}
   	\includegraphics[width=\linewidth]{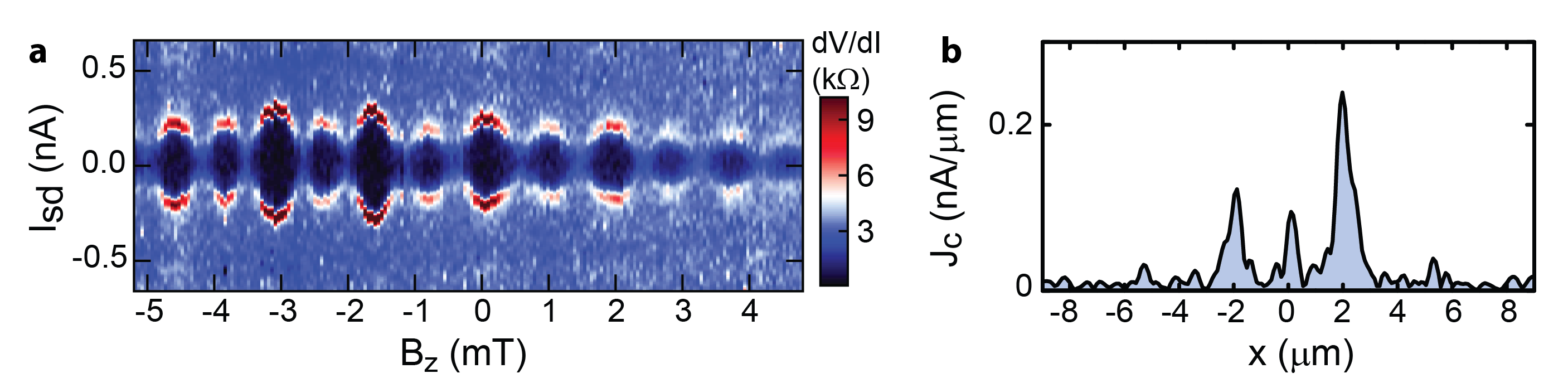}
   	\caption{\label{sup4} \linespread{1.5} \normalsize \textbf{Even-odd effect in the switching current.} (a), $dV/dI_{\text{sd}}$ vs. $B_{\text{z}}$ and $I_{\text{sd}}$ measured on device A at $(V_{\text{bg}}, V_{\text{tg}}) = (-0.8, 4.8)$ V. The data show an even-odd alternation of the switching current amplitude, as in Fig. 5 of the main text, but here measured at different gate settings. We emphasize that this pattern, which effectively doubles the period of the SQI to $2\Phi_0$, is robust. We observed it across a wide range of gate space (see Fig. 2(c) and Fig. 5(a) in the main text), as well as in one other device (see Fig. S12(a)). Based on the Dynes and Fulton approach \cite{01_Dynes1971,02_Hart2014}, for a $2\pi$-periodic current-phase relation the observed pattern translates into a current density profile with a third peak near the device center, as shown in (b). As explained in the main text, this period doubling effect could also result from the fractional Josephson effect \cite{03_Jiang2011, 04_Pikulin2012, 05_San-Jose2012, 06_Dominguez2012}, which is expected to lead to an SQI periodicity of $2\Phi_0$ for 2D TI helical edge modes. }
\end{figure}

\begin{figure}
   	\includegraphics[width=\linewidth]{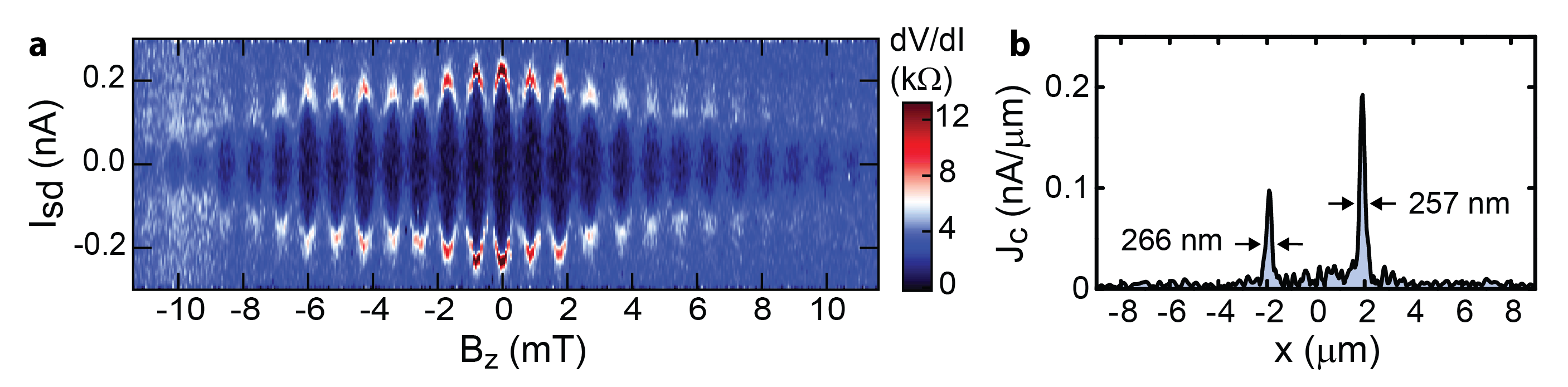}
   	\caption{\label{sup5} \linespread{1.5} \normalsize \textbf{SQI pattern over a large magnetic field range.} (a), $dV/dI_{\text{sd}}$ vs. $B_{\text{z}}$ and $I_{\text{sd}}$ measured on device A at $(V_{\text{bg}}, V_{\text{tg}}) = (-0.4, -0.15)$ V over a larger magnetic field range. The supercurrent oscillations disappear at $\pm 11$ mT, which we attribute to the suppression of superconductivity in the contacts. By resolving more SQI oscillations over a flux range $\Delta \Phi$ corresponding to $\Delta B_{\text{z}} \sim 22$ mT, we can enhance the spatial resolution $(\sim W \Phi_0/\Delta \Phi)$ of the current density profile in (b). The extracted full width at half maximum of the current density peaks (marked by the two pairs of arrows) sets an upper bound of $\sim 260$ nm for the edge mode width. }
\end{figure}

\begin{figure}
   	\includegraphics[width=150mm]{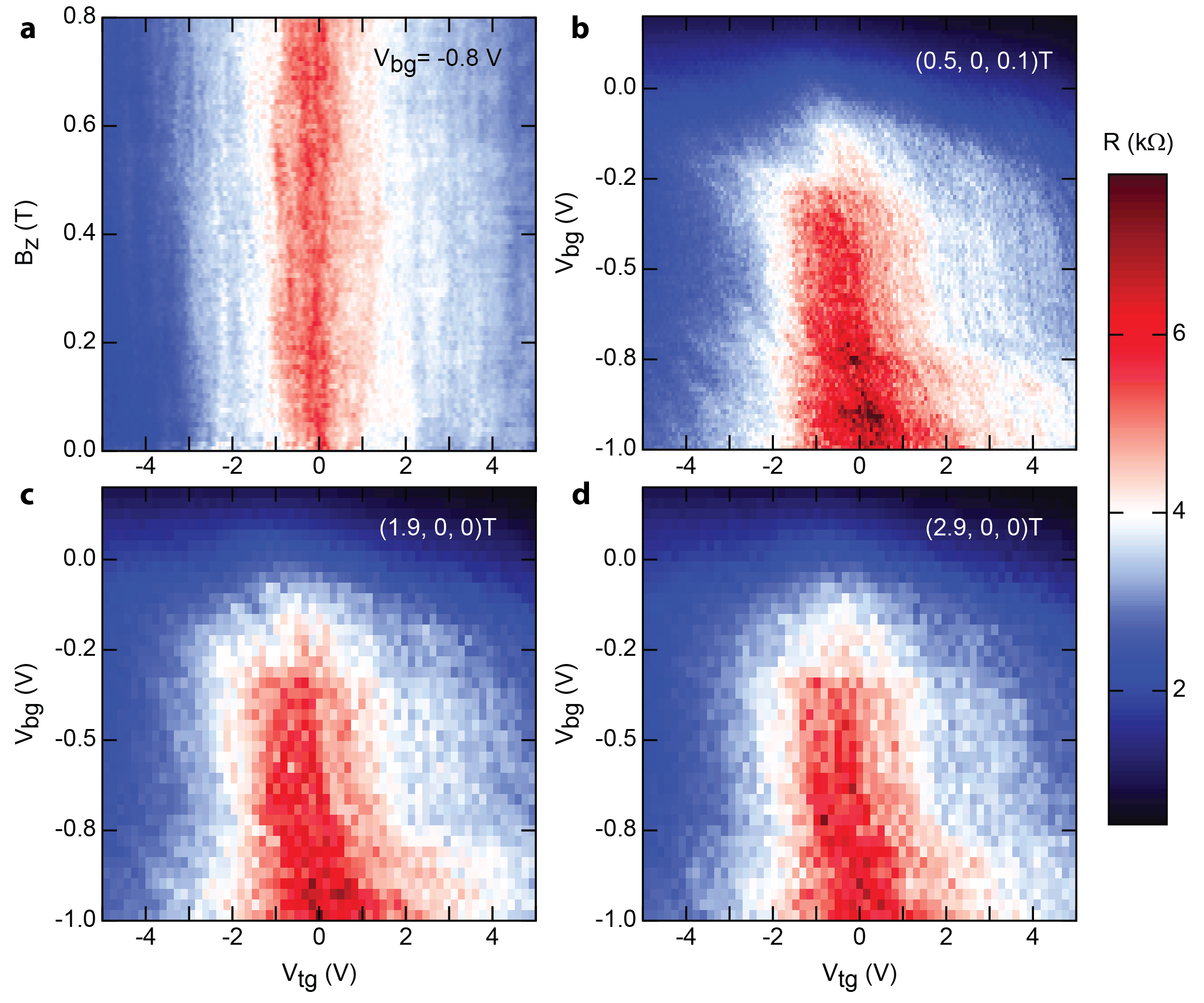}
   	\caption{\label{sup6} \linespread{1.5} \normalsize \textbf{Magnetic-field dependence of the normal state resistance.} No magnetic-field dependence is observed for moderate magnetic fields (device A). (a), $R_{\text{N}}$ vs. $V_{\text{tg}}$ and perpendicular field, $B_{\text{z}}$, for fixed $V_{\text{bg}} = -0.8$ V. No change in $R_{\text{N}}$ is observed when sweeping $B_{\text{z}}$ from 0 T to 0.8 T. (b)-(d), $R_{\text{N}}$ vs. $V_{\text{bg}}$ and $V_{\text{tg}}$ at different magnetic fields, ($B_{\text{x}}, B_{\text{y}}, B_{\text{z}}$) = (0.5, 0, 0.1) T in (b), (1.9, 0, 0) T in (c) and (2.9, 0, 0) T in (d). The $R_{\text{N}}$ phase diagram is almost unchanged between the three fields with increasing in-plane component $B_{\text{x}}$. This is consistent with the lack of in-plane magnetic field dependence reported by Du \textit{et al.} \cite{07_Du2013} and could be due to a very small effective g-factor of the edge modes. To the best of our knowledge, this observation is not well described by existing theoretical models. }
\end{figure}

\begin{figure}
   	\includegraphics[width=140mm]{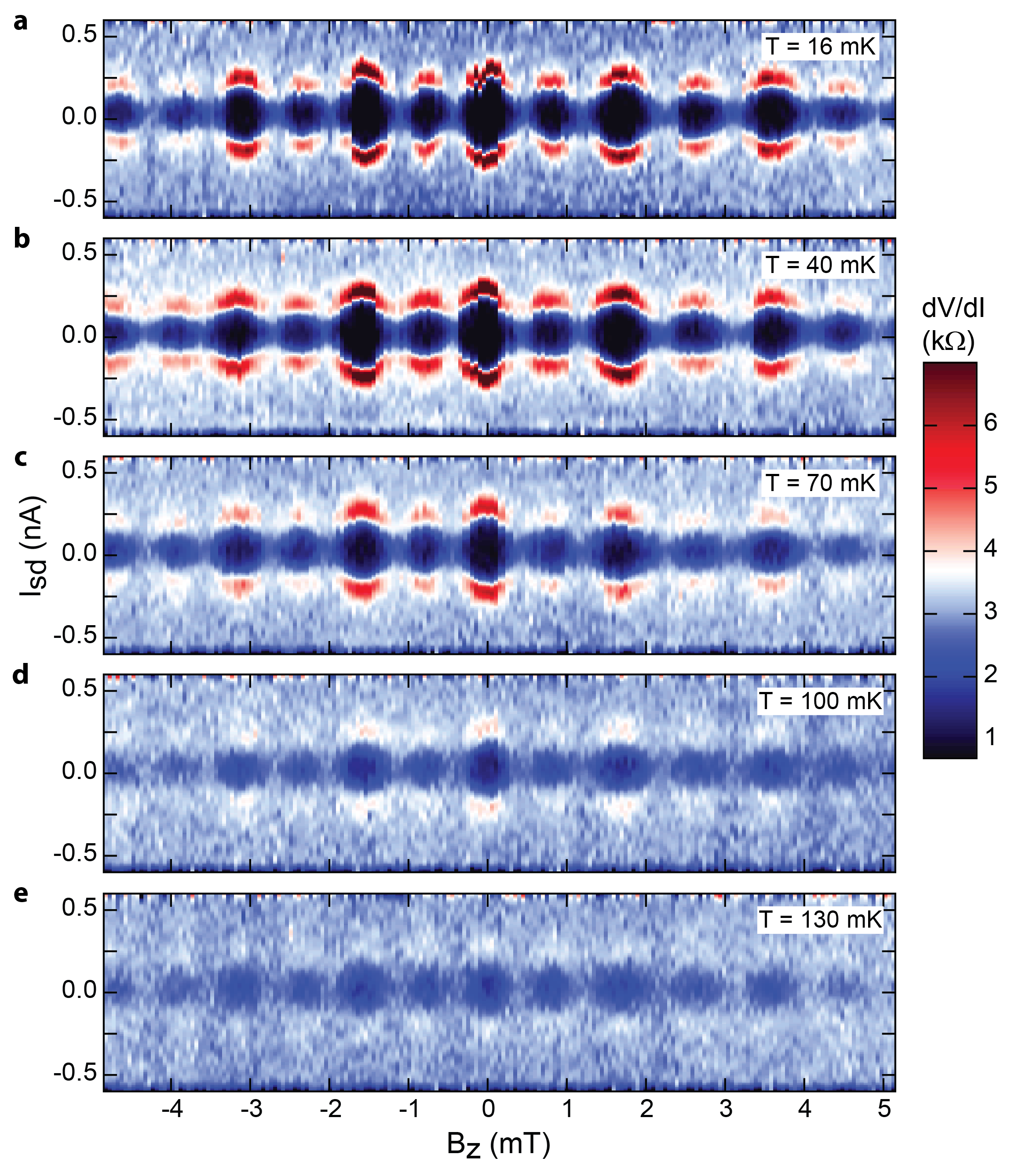}
   	\caption{\label{sup7} \linespread{1.5} \normalsize \textbf{Temperature dependence of the critical current.} (a)-(e), SQI patterns from device A measured for T = 16 mK, 40 mK, 70 mK, 100 mK and 130 mK, respectively. The data sets were measured for fixed $V_{\text{bg}} = -0.8$ V and $V_{\text{tg}} = 5.5$ V, the same values as for Fig. 5(a) in the main text. No difference is observed between 16 mK and 40 mK. The switching current begins to decrease at 70 mK, which likely indicates an effective junction temperature between 40 mK and 70 mK while the mixing chamber is at base temperature. At temperatures above $\sim 130$ mK, the switching currents become small, however the even-odd effect is still observed. }
\end{figure}

\begin{figure}
   	\includegraphics[width=100mm]{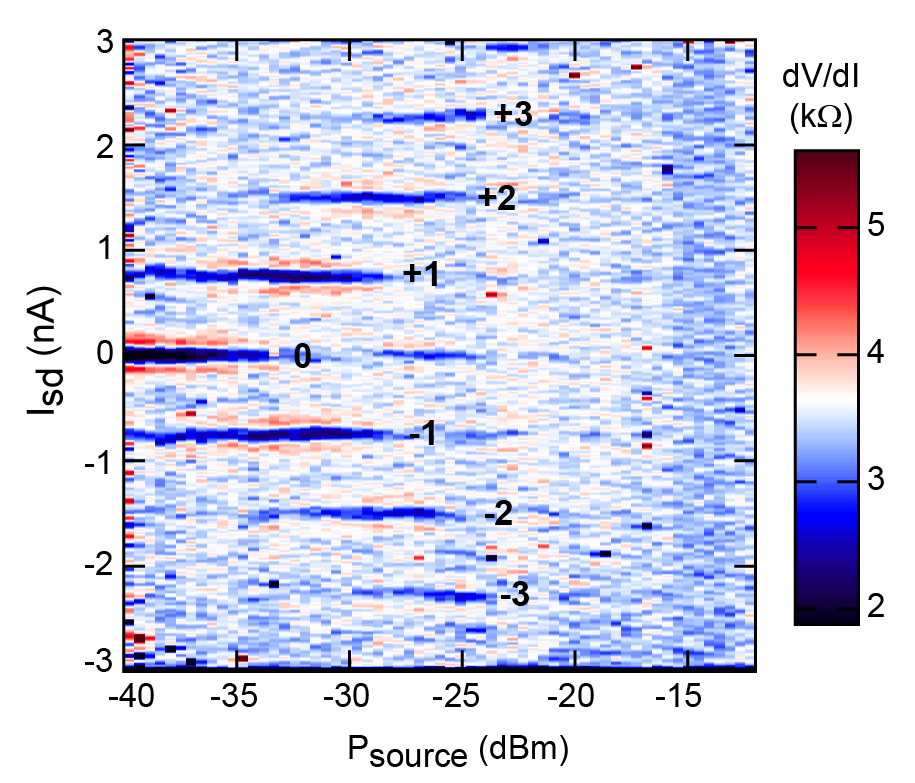}
   	\caption{\label{sup8} \linespread{1.5} \normalsize \textbf{Shapiro steps near the charge neutrality point (CNP).} Shapiro steps \cite{08_Thinkham1996} from device A, at gates settings $V_{\text{bg}} =-0.4$ V and $V_{\text{tg}}=-0.15$ V ($R_{\text{N}} \sim 3500 \ \Omega$). When microwaves of frequency $f_{\text{RF}} =1.288$ GHz are applied, several Shapiro steps develop (numbered $-3$ to $+3$ in the figure). This dataset was measured near the CNP, while the Shapiro steps discussed in the main text (Fig. 3(c)) were measured in the electron regime. A signature of topological superconductivity is the suppression of the odd-number steps due to the fractional Josephson effect \cite{03_Jiang2011, 04_Pikulin2012, 05_San-Jose2012, 06_Dominguez2012}. Here, we do not observe such suppression, which could be due to quasiparticle poisoning. }
\end{figure}

\begin{figure}
   	\includegraphics[width=\linewidth]{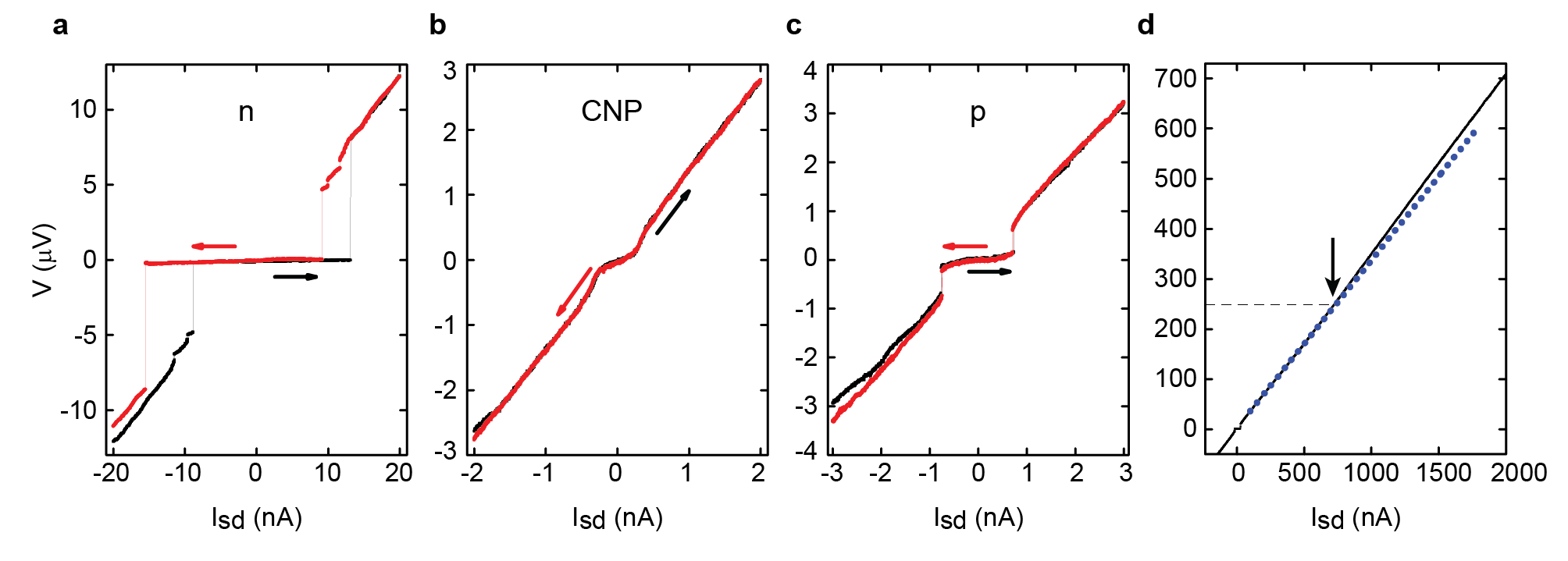}
   	\caption{\label{sup9} \linespread{1.5} \normalsize \textbf{Line-cuts showing representative I-V curves in the three transport regimes.} (a)-(c), I-V curves from device A measured by sweeping $I_{\text{sd}}$ in the two opposite directions at three different gates settings. (a). $(V_{\text{bg}}, V_{\text{tg}}) = (0.1, 5.5)$ V, (b). $(V_{\text{bg}}, V_{\text{tg}}) = (0.1, -1.2)$ V and (c). $(V_{\text{bg}}, V_{\text{tg}}) = (0.1, -5)$ V. The finite slope around $I_{\text{sd}}$ = 0 in (b) and (c) is likely due to temperature broadening effects, which are more effective at low switching currents. (d), I-V curve over a large range of $I_{\text{sd}}$. The blue dotted line is a fit for small positive $I_{\text{sd}}$. The I-V curve deviates from the low-$I_{\text{sd}}$ behaviour above $V \sim 250 \ \mu$V, (see arrow) from which we extract $\Delta \sim 125 \ \mu e$V in the contacts, as expected for our Ti/Al material \cite{09_Doh2005}. Since the I-V slopes for low and high $I_{\text{sd}}$ differ by less than 10\%, we use the slope at low current bias to extract the normal state resistance, $R_{\text{N}}$. }
\end{figure}

\begin{figure}
   	\includegraphics[width=110mm]{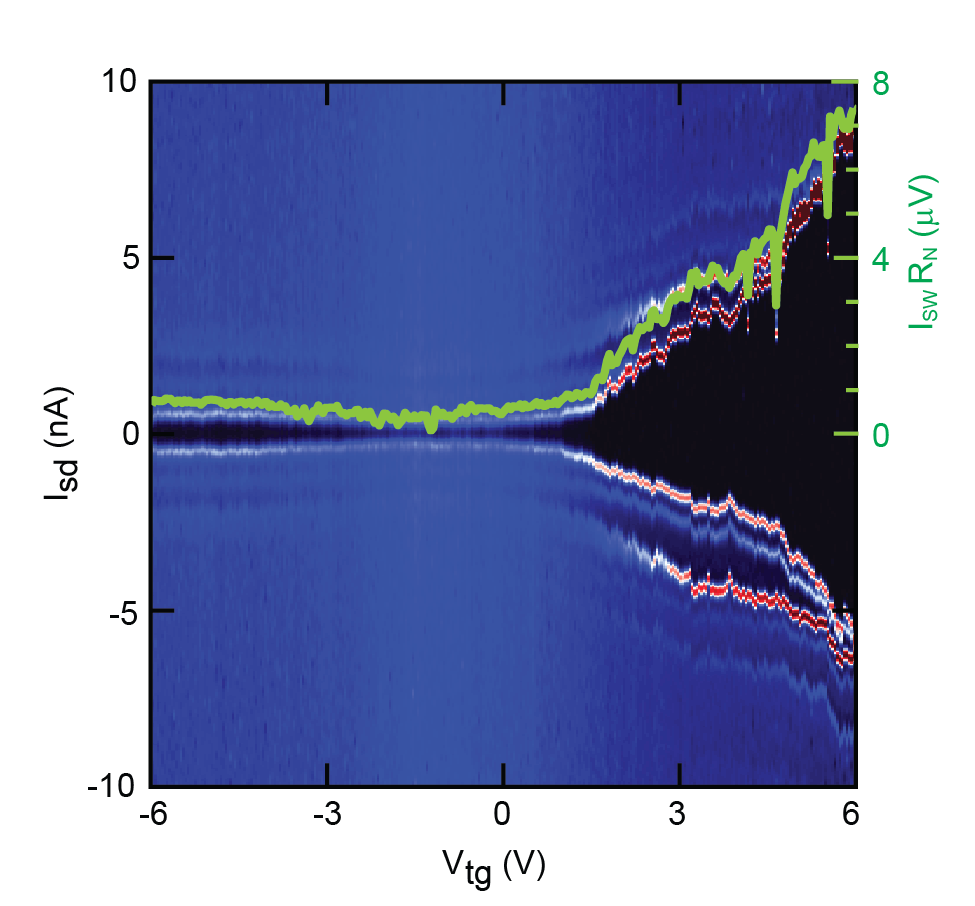}
   	\caption{\label{sup10}\linespread{1.5} \normalsize \textbf{Gate-dependence of the switching current.} Differential resistance as a function of $I_{\text{sd}}$ for device A at $V_{\text{bg}} = 0$ V. 
   	As in Fig. 3(a) in the main text, when sweeping top gate from electron side through charge neutrality point to hole side, the switching currents decrease and then increase again. The green line shows the $I_{\text{SW}}R_{\text{N}}$ product, which closely follows the switching current, $I_{\text{SW}}$. We note that the $I_{\text{SW}}R_{\text{N}}$ product is considerably smaller than the superconducting gap of the Ti/Al contacts, $\Delta/e \sim 125 \ \mu$V. 
   	This may indicate a small induced superconducting gap, $\Delta_{\text{ind}}$, in the InAs/GaSb quantum well or a large suppression of the switching current, $I_{\text{SW}}$, with respect to the critical current, $I_{\text{c}}$, as a result of the electromagnetic environment or thermal activation (note that $k_{\text{B}}T$ is of the order of the $I_{\text{SW}}R_{\text{N}}$ product). }
\end{figure}

\begin{figure}
   	\includegraphics[width=\linewidth]{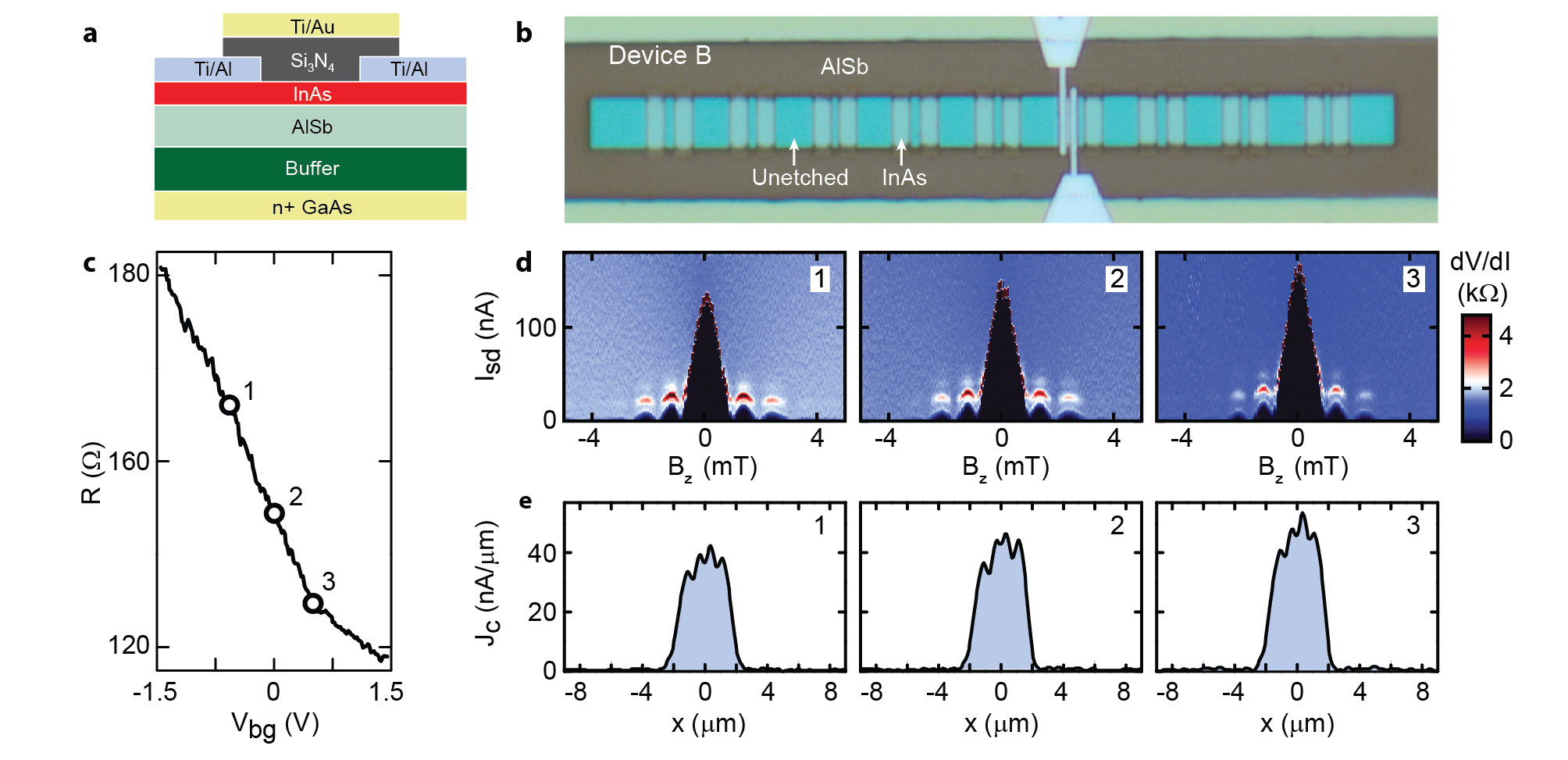}
   	\caption{\label{sup11} \linespread{1.5} \normalsize \textbf{Superconducting and normal transport for an S-InAs-S junction.} Having induced superconductivity in the edge modes (see main text), here we investigated the SQI patterns on a similar junction fabricated on InAs only, where the helical edge modes should be absent. This device (device B) was fabricated by contacting the uncovered InAs layer with Ti/Al. (a), Schematic layout.
   	(b), Optical microscope image of the device before topgate deposition. Device B underwent the same fabrication process as InAs/GaSb device A (both are on the same chip). 
   	(c), Normal state resistance as a function of $V_{\text{bg}}$ measured at $B_{\text{z}}=0$ using a DC excitation current $I_{\text{sd}} = 200$ nA. The top gate was kept floating due to a disconnected bonding pad. (d), $dV/dI$ as a function of $B_{\text{z}}$ at three $V_{\text{bg}}$ values marked by 1, 2 and 3 in (c). 
   	(e), the corresponding current density profiles, showing a uniform current density through the bulk of the InAs layer, as expected for this non-topological junction. We see no evidence for edge modes in InAs, however, completely ruling out the existence of any non-topological edge modes requires gating the device to resistances above $\sim 900 \ \Omega$ (see Fig. S3), which could not be achieved due to the onset of backgate leakage for $V_{\text{bg}} < -1.5$ V. }
\end{figure}

\begin{figure}
   	\includegraphics[width=155mm]{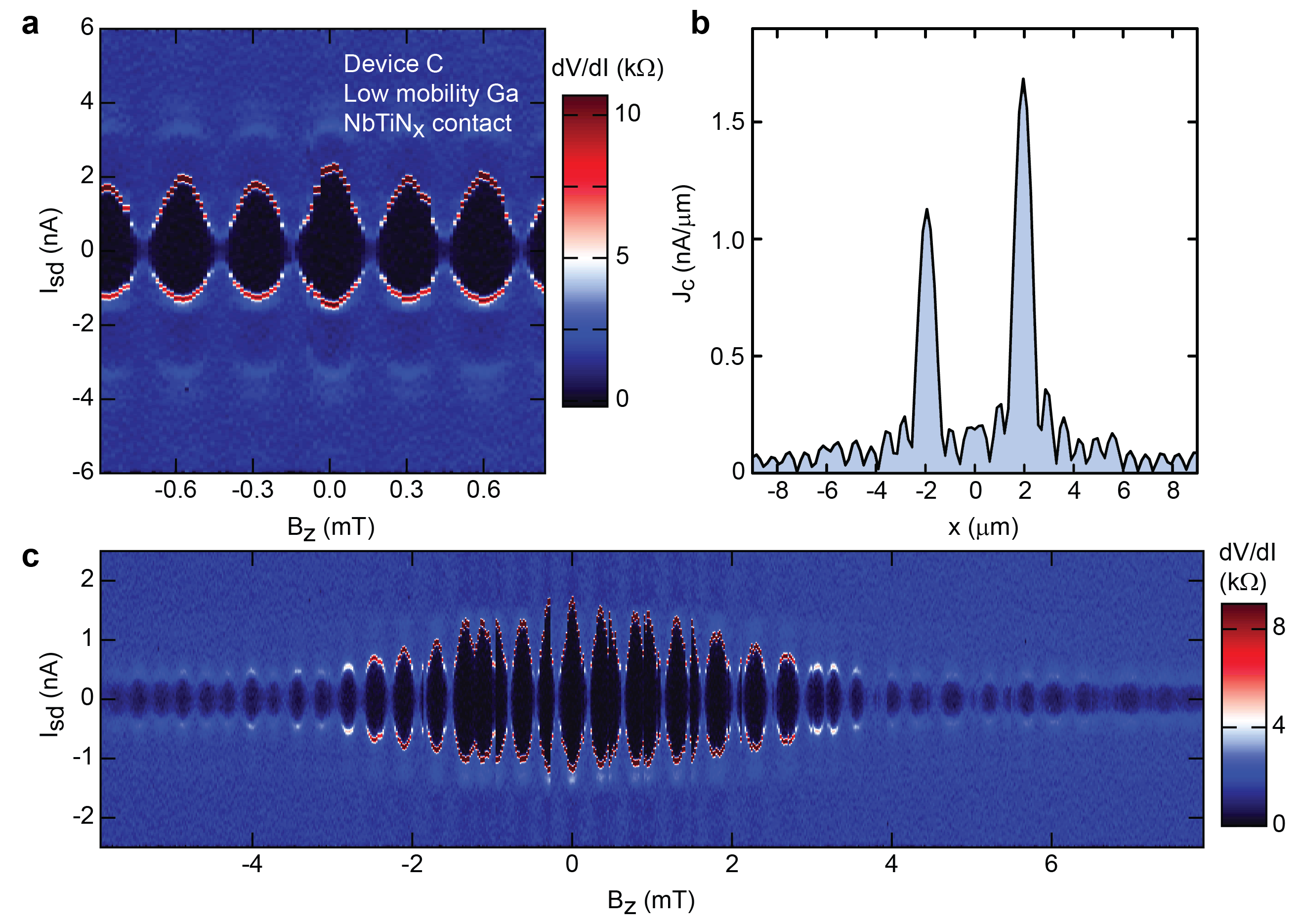}
   	\caption{\label{sup12}\linespread{1.5} \normalsize \textbf{SQI patterns for an S-InAs/GaSb-S junction based on InAs/GaSb grown using a lower mobility Ga source and contacted with $\text{NbTiN}_{\text{x}}$.} This device (device C), was fabricated using an InAs/GaSb quantum well structure grown with a lower mobility Ga source, which suppresses the residual bulk conductivity \cite{10_Charpentier2013}. Differently from devices A and B, the superconducting contacts are made from 200 nm-thick sputtered $\text{NbTiN}_{\text{x}}$, and have a width of 1 $\mu$m. (a), Without any gating, we observe a SQUID-like SQI pattern as a function of $B_{\text{z}}$ (for $R_{\text{N}} \sim 1600 \ \Omega$), which corresponds to edge-mode dominated superconducting transport. By analogy with device A, this suggests that the Fermi level resides in the bulk gap (near the CNP). Similar results were also observed in another similar device (device D, not shown). Further, we note the presence of an even-odd effect, similar to that observed in device A (see Fig. 5 and Fig. S4(a)). (b), The corresponding current density profile extracted using a $2\pi$-periodic current-phase relation. (c), SQUID-like pattern for a larger range in $B_{\text{z}}$. For $B_{\text{z}} > -1.8$ mT the pattern shows switches along the $B_{\text{z}}$-axis, presumably due to flux depinning in the $\text{NbTiN}_{\text{x}}$ contacts (the 200 nm contact thickness is below the London penetration depth of $\sim 250$ nm \cite{11_Yu2005} of $\text{NbTiN}_{\text{x}}$). The switching behaviour is hysteretic in the field-sweep direction (not shown), which is consistent with flux-depinning. Despite the switches, it is clear that the oscillations in (c) show very little attenuation at larger magnetic field amplitudes, consistent with a SQUID behaviour. }
\end{figure}

\begin{figure}
   	\includegraphics[width=130mm]{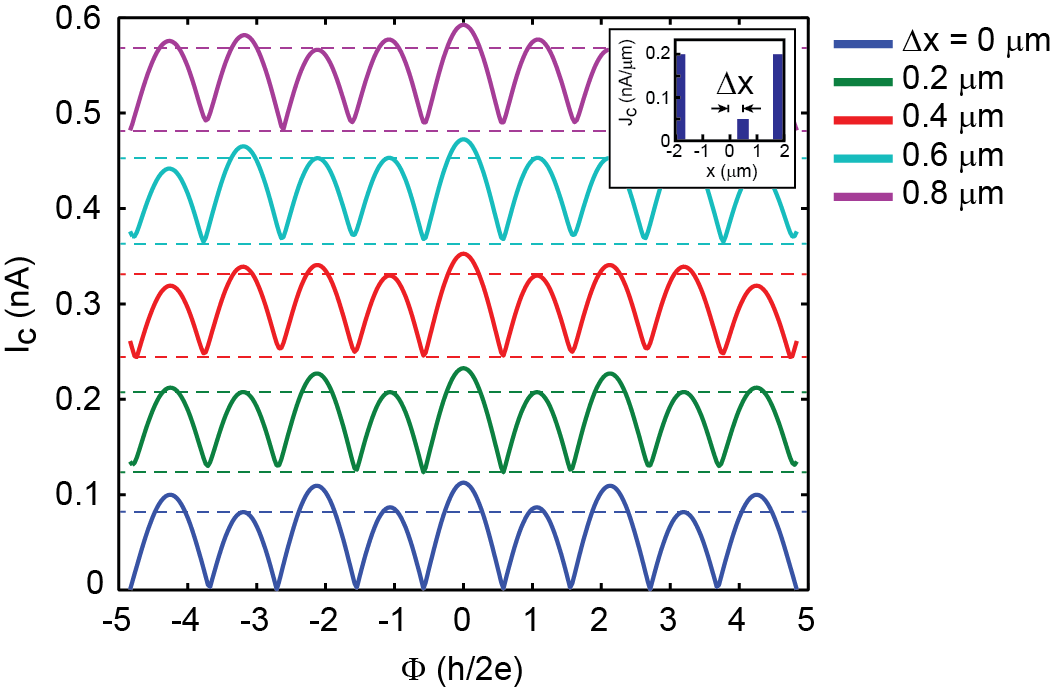}
   	\caption{\label{sup13}\linespread{1.5} \normalsize \textbf{Simulated SQI patterns.} Simulation results for a SQUID, assuming a third transport channel located at a distance $\Delta x$ from the center. We use a conventional, $2\pi$-periodic current-phase relation. The inset shows the current distribution for which the SQI patterns are calculated. The edge and center modes each have a width of 0.25 $\mu $m. The central mode has a critical current density 4 times smaller than that of the edges, in order to fit the measured data. An even-odd effect similar to experimental observations (see Fig. 5(a) in the main text, Fig. S4(a), and Fig. S12(a)) is obtained for $\Delta x = 0 \ \mu$m (third mode exactly in the middle). The effect is not seen in simulations where $\Delta x > 0.2 \ \mu$m. This restricts the position of a possible third channel within $\sim 200$ nm from the center. Curves are offset from each other by 0.12 nA for clarity.}
\end{figure}

\clearpage


\bibliographystyleS{apsrev4-1}

%

%% file: Manuscript.bbl
\begin{thebibliography}{29}%
\makeatletter
\providecommand \@ifxundefined [1]{%
 \@ifx{#1\undefined}
}%
\providecommand \@ifnum [1]{%
 \ifnum #1\expandafter \@firstoftwo
 \else \expandafter \@secondoftwo
 \fi
}%
\providecommand \@ifx [1]{%
 \ifx #1\expandafter \@firstoftwo
 \else \expandafter \@secondoftwo
 \fi
}%
\providecommand \natexlab [1]{#1}%
\providecommand \enquote  [1]{``#1''}%
\providecommand \bibnamefont  [1]{#1}%
\providecommand \bibfnamefont [1]{#1}%
\providecommand \citenamefont [1]{#1}%
\providecommand \href@noop [0]{\@secondoftwo}%
\providecommand \href [0]{\begingroup \@sanitize@url \@href}%
\providecommand \@href[1]{\@@startlink{#1}\@@href}%
\providecommand \@@href[1]{\endgroup#1\@@endlink}%
\providecommand \@sanitize@url [0]{\catcode `\\12\catcode `\$12\catcode
  `\&12\catcode `\#12\catcode `\^12\catcode `\_12\catcode `\%12\relax}%
\providecommand \@@startlink[1]{}%
\providecommand \@@endlink[0]{}%
\providecommand \url  [0]{\begingroup\@sanitize@url \@url }%
\providecommand \@url [1]{\endgroup\@href {#1}{\urlprefix }}%
\providecommand \urlprefix  [0]{URL }%
\providecommand \Eprint [0]{\href }%
\providecommand \doibase [0]{http://dx.doi.org/}%
\providecommand \selectlanguage [0]{\@gobble}%
\providecommand \bibinfo  [0]{\@secondoftwo}%
\providecommand \bibfield  [0]{\@secondoftwo}%
\providecommand \translation [1]{[#1]}%
\providecommand \BibitemOpen [0]{}%
\providecommand \bibitemStop [0]{}%
\providecommand \bibitemNoStop [0]{.\EOS\space}%
\providecommand \EOS [0]{\spacefactor3000\relax}%
\providecommand \BibitemShut  [1]{\csname bibitem#1\endcsname}%
\let\auto@bib@innerbib\@empty
\bibitem [{\citenamefont {Hasan}\ and\ \citenamefont
  {Kane}(2010)}]{01_Hasan2010}%
  \BibitemOpen
  \bibfield  {author} {\bibinfo {author} {\bibfnamefont {M.~Z.}\ \bibnamefont
  {Hasan}}\ and\ \bibinfo {author} {\bibfnamefont {C.~L.}\ \bibnamefont
  {Kane}},\ }\href {\doibase 10.1103/RevModPhys.82.3045} {\bibfield  {journal}
  {\bibinfo  {journal} {Rev. Mod. Phys.}\ }\textbf {\bibinfo {volume} {82}},\
  \bibinfo {pages} {3045} (\bibinfo {year} {2010})}\BibitemShut {NoStop}%
\bibitem [{\citenamefont {Qi}\ and\ \citenamefont {Zhang}(2011)}]{02_Qi2011}%
  \BibitemOpen
  \bibfield  {author} {\bibinfo {author} {\bibfnamefont {X.-L.}\ \bibnamefont
  {Qi}}\ and\ \bibinfo {author} {\bibfnamefont {S.-C.}\ \bibnamefont {Zhang}},\
  }\href {\doibase 10.1103/RevModPhys.83.1057} {\bibfield  {journal} {\bibinfo
  {journal} {Rev. Mod. Phys.}\ }\textbf {\bibinfo {volume} {83}},\ \bibinfo
  {pages} {1057} (\bibinfo {year} {2011})}\BibitemShut {NoStop}%
\bibitem [{\citenamefont {Nayak}\ \emph {et~al.}(2008)\citenamefont {Nayak},
  \citenamefont {Stern}, \citenamefont {Freedman},\ and\ \citenamefont {{Das
  Sarma}}}]{03_Nayak2008}%
  \BibitemOpen
  \bibfield  {author} {\bibinfo {author} {\bibfnamefont {C.}~\bibnamefont
  {Nayak}}, \bibinfo {author} {\bibfnamefont {A.}~\bibnamefont {Stern}},
  \bibinfo {author} {\bibfnamefont {M.}~\bibnamefont {Freedman}}, \ and\
  \bibinfo {author} {\bibfnamefont {S.}~\bibnamefont {{Das Sarma}}},\ }\href
  {\doibase 10.1103/RevModPhys.80.1083} {\bibfield  {journal} {\bibinfo
  {journal} {Rev. Mod. Phys.}\ }\textbf {\bibinfo {volume} {80}},\ \bibinfo
  {pages} {1083} (\bibinfo {year} {2008})}\BibitemShut {NoStop}%
\bibitem [{\citenamefont {Read}(2012)}]{04_Read2012}%
  \BibitemOpen
  \bibfield  {author} {\bibinfo {author} {\bibfnamefont {N.}~\bibnamefont
  {Read}},\ }\href {\doibase 10.1063/PT.3.1641} {\bibfield  {journal} {\bibinfo
   {journal} {Phys. Today}\ }\textbf {\bibinfo {volume} {65}},\ \bibinfo
  {pages} {38} (\bibinfo {year} {2012})}\BibitemShut {NoStop}%
\bibitem [{\citenamefont {Kane}\ and\ \citenamefont
  {Mele}(2005)}]{05_Kane2005}%
  \BibitemOpen
  \bibfield  {author} {\bibinfo {author} {\bibfnamefont {C.~L.}\ \bibnamefont
  {Kane}}\ and\ \bibinfo {author} {\bibfnamefont {E.~J.}\ \bibnamefont
  {Mele}},\ }\href {\doibase 10.1103/PhysRevLett.95.146802} {\bibfield
  {journal} {\bibinfo  {journal} {Phys. Rev. Lett.}\ }\textbf {\bibinfo
  {volume} {95}},\ \bibinfo {pages} {146802} (\bibinfo {year}
  {2005})}\BibitemShut {NoStop}%
\bibitem [{\citenamefont {Bernevig}\ and\ \citenamefont
  {Zhang}(2006)}]{06_Bernevig2006}%
  \BibitemOpen
  \bibfield  {author} {\bibinfo {author} {\bibfnamefont {B.~A.}\ \bibnamefont
  {Bernevig}}\ and\ \bibinfo {author} {\bibfnamefont {S.-C.}\ \bibnamefont
  {Zhang}},\ }\href {\doibase 10.1103/PhysRevLett.96.106802} {\bibfield
  {journal} {\bibinfo  {journal} {Phys. Rev. Lett.}\ }\textbf {\bibinfo
  {volume} {96}},\ \bibinfo {pages} {106802} (\bibinfo {year}
  {2006})}\BibitemShut {NoStop}%
\bibitem [{\citenamefont {Bernevig}\ \emph {et~al.}(2006)\citenamefont
  {Bernevig}, \citenamefont {Hughes},\ and\ \citenamefont
  {Zhang}}]{07_Bernevig2006a}%
  \BibitemOpen
  \bibfield  {author} {\bibinfo {author} {\bibfnamefont {B.~A.}\ \bibnamefont
  {Bernevig}}, \bibinfo {author} {\bibfnamefont {T.~L.}\ \bibnamefont
  {Hughes}}, \ and\ \bibinfo {author} {\bibfnamefont {S.-C.}\ \bibnamefont
  {Zhang}},\ }\href {\doibase 10.1126/science.1133734} {\bibfield  {journal}
  {\bibinfo  {journal} {Science}\ }\textbf {\bibinfo {volume} {314}},\ \bibinfo
  {pages} {1757} (\bibinfo {year} {2006})}\BibitemShut {NoStop}%
\bibitem [{\citenamefont {K\"{o}nig}\ \emph {et~al.}(2007)\citenamefont
  {K\"{o}nig}, \citenamefont {Wiedmann}, \citenamefont {Br\"{u}ne},
  \citenamefont {Roth}, \citenamefont {Buhmann}, \citenamefont {Molenkamp},
  \citenamefont {Qi},\ and\ \citenamefont {Zhang}}]{08_Konig2007}%
  \BibitemOpen
  \bibfield  {author} {\bibinfo {author} {\bibfnamefont {M.}~\bibnamefont
  {K\"{o}nig}}, \bibinfo {author} {\bibfnamefont {S.}~\bibnamefont {Wiedmann}},
  \bibinfo {author} {\bibfnamefont {C.}~\bibnamefont {Br\"{u}ne}}, \bibinfo
  {author} {\bibfnamefont {A.}~\bibnamefont {Roth}}, \bibinfo {author}
  {\bibfnamefont {H.}~\bibnamefont {Buhmann}}, \bibinfo {author} {\bibfnamefont
  {L.~W.}\ \bibnamefont {Molenkamp}}, \bibinfo {author} {\bibfnamefont {X.-L.}\
  \bibnamefont {Qi}}, \ and\ \bibinfo {author} {\bibfnamefont {S.-C.}\
  \bibnamefont {Zhang}},\ }\href {\doibase 10.1126/science.1148047} {\bibfield
  {journal} {\bibinfo  {journal} {Science}\ }\textbf {\bibinfo {volume}
  {318}},\ \bibinfo {pages} {766} (\bibinfo {year} {2007})}\BibitemShut
  {NoStop}%
\bibitem [{\citenamefont {Liu}\ \emph {et~al.}(2008)\citenamefont {Liu},
  \citenamefont {Hughes}, \citenamefont {Qi}, \citenamefont {Wang},\ and\
  \citenamefont {Zhang}}]{09_Liu2008}%
  \BibitemOpen
  \bibfield  {author} {\bibinfo {author} {\bibfnamefont {C.}~\bibnamefont
  {Liu}}, \bibinfo {author} {\bibfnamefont {T.}~\bibnamefont {Hughes}},
  \bibinfo {author} {\bibfnamefont {X.-L.}\ \bibnamefont {Qi}}, \bibinfo
  {author} {\bibfnamefont {K.}~\bibnamefont {Wang}}, \ and\ \bibinfo {author}
  {\bibfnamefont {S.-C.}\ \bibnamefont {Zhang}},\ }\href {\doibase
  10.1103/PhysRevLett.100.236601} {\bibfield  {journal} {\bibinfo  {journal}
  {Phys. Rev. Lett.}\ }\textbf {\bibinfo {volume} {100}},\ \bibinfo {pages}
  {236601} (\bibinfo {year} {2008})}\BibitemShut {NoStop}%
\bibitem [{\citenamefont {Du}\ \emph {et~al.}()\citenamefont {Du},
  \citenamefont {Knez}, \citenamefont {Sullivan},\ and\ \citenamefont
  {Du}}]{10_Du2013}%
  \BibitemOpen
  \bibfield  {author} {\bibinfo {author} {\bibfnamefont {L.}~\bibnamefont
  {Du}}, \bibinfo {author} {\bibfnamefont {I.}~\bibnamefont {Knez}}, \bibinfo
  {author} {\bibfnamefont {G.}~\bibnamefont {Sullivan}}, \ and\ \bibinfo
  {author} {\bibfnamefont {R.-R.}\ \bibnamefont {Du}},\ }\href
  {http://arxiv.org/abs/1306.1925} {\ }\Eprint {http://arxiv.org/abs/1306.1925}
  {arXiv:1306.1925} \BibitemShut {NoStop}%
\bibitem [{\citenamefont {Sasaki}\ \emph {et~al.}(2011)\citenamefont {Sasaki},
  \citenamefont {Kriener}, \citenamefont {Segawa}, \citenamefont {Yada},
  \citenamefont {Tanaka}, \citenamefont {Sato},\ and\ \citenamefont
  {Ando}}]{11_Sasaki2011}%
  \BibitemOpen
  \bibfield  {author} {\bibinfo {author} {\bibfnamefont {S.}~\bibnamefont
  {Sasaki}}, \bibinfo {author} {\bibfnamefont {M.}~\bibnamefont {Kriener}},
  \bibinfo {author} {\bibfnamefont {K.}~\bibnamefont {Segawa}}, \bibinfo
  {author} {\bibfnamefont {K.}~\bibnamefont {Yada}}, \bibinfo {author}
  {\bibfnamefont {Y.}~\bibnamefont {Tanaka}}, \bibinfo {author} {\bibfnamefont
  {M.}~\bibnamefont {Sato}}, \ and\ \bibinfo {author} {\bibfnamefont
  {Y.}~\bibnamefont {Ando}},\ }\href {\doibase 10.1103/PhysRevLett.107.217001}
  {\bibfield  {journal} {\bibinfo  {journal} {Phys. Rev. Lett.}\ }\textbf
  {\bibinfo {volume} {107}},\ \bibinfo {pages} {217001} (\bibinfo {year}
  {2011})}\BibitemShut {NoStop}%
\bibitem [{\citenamefont {Mourik}\ \emph {et~al.}(2012)\citenamefont {Mourik},
  \citenamefont {Zuo}, \citenamefont {Frolov}, \citenamefont {Plissard},
  \citenamefont {Bakkers},\ and\ \citenamefont {Kouwenhoven}}]{12_Mourik2012}%
  \BibitemOpen
  \bibfield  {author} {\bibinfo {author} {\bibfnamefont {V.}~\bibnamefont
  {Mourik}}, \bibinfo {author} {\bibfnamefont {K.}~\bibnamefont {Zuo}},
  \bibinfo {author} {\bibfnamefont {S.~M.}\ \bibnamefont {Frolov}}, \bibinfo
  {author} {\bibfnamefont {S.~R.}\ \bibnamefont {Plissard}}, \bibinfo {author}
  {\bibfnamefont {E.~P. a.~M.}\ \bibnamefont {Bakkers}}, \ and\ \bibinfo
  {author} {\bibfnamefont {L.~P.}\ \bibnamefont {Kouwenhoven}},\ }\href
  {\doibase 10.1126/science.1222360} {\bibfield  {journal} {\bibinfo  {journal}
  {Science}\ }\textbf {\bibinfo {volume} {336}},\ \bibinfo {pages} {1003}
  (\bibinfo {year} {2012})}\BibitemShut {NoStop}%
\bibitem [{\citenamefont {Das}\ \emph {et~al.}(2012)\citenamefont {Das},
  \citenamefont {Ronen}, \citenamefont {Most}, \citenamefont {Oreg},
  \citenamefont {Heiblum},\ and\ \citenamefont {Shtrikman}}]{13_Das2012}%
  \BibitemOpen
  \bibfield  {author} {\bibinfo {author} {\bibfnamefont {A.}~\bibnamefont
  {Das}}, \bibinfo {author} {\bibfnamefont {Y.}~\bibnamefont {Ronen}}, \bibinfo
  {author} {\bibfnamefont {Y.}~\bibnamefont {Most}}, \bibinfo {author}
  {\bibfnamefont {Y.}~\bibnamefont {Oreg}}, \bibinfo {author} {\bibfnamefont
  {M.}~\bibnamefont {Heiblum}}, \ and\ \bibinfo {author} {\bibfnamefont
  {H.}~\bibnamefont {Shtrikman}},\ }\href {\doibase 10.1038/nphys2479}
  {\bibfield  {journal} {\bibinfo  {journal} {Nat. Phys.}\ }\textbf {\bibinfo
  {volume} {8}},\ \bibinfo {pages} {887} (\bibinfo {year} {2012})}\BibitemShut
  {NoStop}%
\bibitem [{\citenamefont {Deng}\ \emph {et~al.}(2012)\citenamefont {Deng},
  \citenamefont {Yu}, \citenamefont {Huang}, \citenamefont {Larsson},
  \citenamefont {Caroff},\ and\ \citenamefont {Xu}}]{14_Deng2012}%
  \BibitemOpen
  \bibfield  {author} {\bibinfo {author} {\bibfnamefont {M.~T.}\ \bibnamefont
  {Deng}}, \bibinfo {author} {\bibfnamefont {C.~L.}\ \bibnamefont {Yu}},
  \bibinfo {author} {\bibfnamefont {G.~Y.}\ \bibnamefont {Huang}}, \bibinfo
  {author} {\bibfnamefont {M.}~\bibnamefont {Larsson}}, \bibinfo {author}
  {\bibfnamefont {P.}~\bibnamefont {Caroff}}, \ and\ \bibinfo {author}
  {\bibfnamefont {H.~Q.}\ \bibnamefont {Xu}},\ }\href {\doibase
  10.1021/nl303758w} {\bibfield  {journal} {\bibinfo  {journal} {Nano Lett.}\
  }\textbf {\bibinfo {volume} {12}},\ \bibinfo {pages} {6414} (\bibinfo {year}
  {2012})}\BibitemShut {NoStop}%
\bibitem [{\citenamefont {Churchill}\ \emph {et~al.}(2013)\citenamefont
  {Churchill}, \citenamefont {Fatemi}, \citenamefont {Grove-Rasmussen},
  \citenamefont {Deng}, \citenamefont {Caroff}, \citenamefont {Xu},\ and\
  \citenamefont {Marcus}}]{15_Churchill2013}%
  \BibitemOpen
  \bibfield  {author} {\bibinfo {author} {\bibfnamefont {H.~O.~H.}\
  \bibnamefont {Churchill}}, \bibinfo {author} {\bibfnamefont {V.}~\bibnamefont
  {Fatemi}}, \bibinfo {author} {\bibfnamefont {K.}~\bibnamefont
  {Grove-Rasmussen}}, \bibinfo {author} {\bibfnamefont {M.~T.}\ \bibnamefont
  {Deng}}, \bibinfo {author} {\bibfnamefont {P.}~\bibnamefont {Caroff}},
  \bibinfo {author} {\bibfnamefont {H.~Q.}\ \bibnamefont {Xu}}, \ and\ \bibinfo
  {author} {\bibfnamefont {C.~M.}\ \bibnamefont {Marcus}},\ }\href {\doibase
  10.1103/PhysRevB.87.241401} {\bibfield  {journal} {\bibinfo  {journal} {Phys.
  Rev. B}\ }\textbf {\bibinfo {volume} {87}},\ \bibinfo {pages} {241401}
  (\bibinfo {year} {2013})}\BibitemShut {NoStop}%
\bibitem [{\citenamefont {Knez}\ \emph {et~al.}(2011)\citenamefont {Knez},
  \citenamefont {Du},\ and\ \citenamefont {Sullivan}}]{16_Knez2011a}%
  \BibitemOpen
  \bibfield  {author} {\bibinfo {author} {\bibfnamefont {I.}~\bibnamefont
  {Knez}}, \bibinfo {author} {\bibfnamefont {R.-R.}\ \bibnamefont {Du}}, \ and\
  \bibinfo {author} {\bibfnamefont {G.}~\bibnamefont {Sullivan}},\ }\href
  {\doibase 10.1103/PhysRevLett.107.136603} {\bibfield  {journal} {\bibinfo
  {journal} {Phys. Rev. Lett.}\ }\textbf {\bibinfo {volume} {107}},\ \bibinfo
  {pages} {136603} (\bibinfo {year} {2011})}\BibitemShut {NoStop}%
\bibitem [{\citenamefont {Nowack}\ \emph {et~al.}(2013)\citenamefont {Nowack},
  \citenamefont {Spanton}, \citenamefont {Baenninger}, \citenamefont
  {K\"{o}nig}, \citenamefont {Kirtley}, \citenamefont {Kalisky}, \citenamefont
  {Ames}, \citenamefont {Leubner}, \citenamefont {Br\"{u}ne}, \citenamefont
  {Buhmann}, \citenamefont {Molenkamp}, \citenamefont {Goldhaber-Gordon},\ and\
  \citenamefont {Moler}}]{17_Nowack2013}%
  \BibitemOpen
  \bibfield  {author} {\bibinfo {author} {\bibfnamefont {K.~C.}\ \bibnamefont
  {Nowack}}, \bibinfo {author} {\bibfnamefont {E.~M.}\ \bibnamefont {Spanton}},
  \bibinfo {author} {\bibfnamefont {M.}~\bibnamefont {Baenninger}}, \bibinfo
  {author} {\bibfnamefont {M.}~\bibnamefont {K\"{o}nig}}, \bibinfo {author}
  {\bibfnamefont {J.~R.}\ \bibnamefont {Kirtley}}, \bibinfo {author}
  {\bibfnamefont {B.}~\bibnamefont {Kalisky}}, \bibinfo {author} {\bibfnamefont
  {C.}~\bibnamefont {Ames}}, \bibinfo {author} {\bibfnamefont {P.}~\bibnamefont
  {Leubner}}, \bibinfo {author} {\bibfnamefont {C.}~\bibnamefont {Br\"{u}ne}},
  \bibinfo {author} {\bibfnamefont {H.}~\bibnamefont {Buhmann}}, \bibinfo
  {author} {\bibfnamefont {L.~W.}\ \bibnamefont {Molenkamp}}, \bibinfo {author}
  {\bibfnamefont {D.}~\bibnamefont {Goldhaber-Gordon}}, \ and\ \bibinfo
  {author} {\bibfnamefont {K.~A.}\ \bibnamefont {Moler}},\ }\href {\doibase
  10.1038/nmat3682} {\bibfield  {journal} {\bibinfo  {journal} {Nat. Mater.}\
  }\textbf {\bibinfo {volume} {12}},\ \bibinfo {pages} {787} (\bibinfo {year}
  {2013})}\BibitemShut {NoStop}%
\bibitem [{\citenamefont {Spanton}\ \emph {et~al.}(2014)\citenamefont
  {Spanton}, \citenamefont {Nowack}, \citenamefont {Du}, \citenamefont
  {Sullivan}, \citenamefont {Du},\ and\ \citenamefont
  {Moler}}]{18_Spanton2014}%
  \BibitemOpen
  \bibfield  {author} {\bibinfo {author} {\bibfnamefont {E.~M.}\ \bibnamefont
  {Spanton}}, \bibinfo {author} {\bibfnamefont {K.~C.}\ \bibnamefont {Nowack}},
  \bibinfo {author} {\bibfnamefont {L.}~\bibnamefont {Du}}, \bibinfo {author}
  {\bibfnamefont {G.}~\bibnamefont {Sullivan}}, \bibinfo {author}
  {\bibfnamefont {R.-R.}\ \bibnamefont {Du}}, \ and\ \bibinfo {author}
  {\bibfnamefont {K.~A.}\ \bibnamefont {Moler}},\ }\href {\doibase
  10.1103/PhysRevLett.113.026804} {\bibfield  {journal} {\bibinfo  {journal}
  {Phys. Rev. Lett.}\ }\textbf {\bibinfo {volume} {113}},\ \bibinfo {pages}
  {026804} (\bibinfo {year} {2014})}\BibitemShut {NoStop}%
\bibitem [{\citenamefont {Knez}\ \emph {et~al.}(2012)\citenamefont {Knez},
  \citenamefont {Du},\ and\ \citenamefont {Sullivan}}]{19_Knez2012}%
  \BibitemOpen
  \bibfield  {author} {\bibinfo {author} {\bibfnamefont {I.}~\bibnamefont
  {Knez}}, \bibinfo {author} {\bibfnamefont {R.-R.}\ \bibnamefont {Du}}, \ and\
  \bibinfo {author} {\bibfnamefont {G.}~\bibnamefont {Sullivan}},\ }\href
  {\doibase 10.1103/PhysRevLett.109.186603} {\bibfield  {journal} {\bibinfo
  {journal} {Phys. Rev. Lett.}\ }\textbf {\bibinfo {volume} {109}},\ \bibinfo
  {pages} {186603} (\bibinfo {year} {2012})}\BibitemShut {NoStop}%
\bibitem [{\citenamefont {Oostinga}\ \emph {et~al.}(2013)\citenamefont
  {Oostinga}, \citenamefont {Maier}, \citenamefont {Sch\"{u}ffelgen},
  \citenamefont {Knott}, \citenamefont {Ames}, \citenamefont {Br\"{u}ne},
  \citenamefont {Tkachov}, \citenamefont {Buhmann},\ and\ \citenamefont
  {Molenkamp}}]{20_Oostinga2013}%
  \BibitemOpen
  \bibfield  {author} {\bibinfo {author} {\bibfnamefont {J.~B.}\ \bibnamefont
  {Oostinga}}, \bibinfo {author} {\bibfnamefont {L.}~\bibnamefont {Maier}},
  \bibinfo {author} {\bibfnamefont {P.}~\bibnamefont {Sch\"{u}ffelgen}},
  \bibinfo {author} {\bibfnamefont {D.}~\bibnamefont {Knott}}, \bibinfo
  {author} {\bibfnamefont {C.}~\bibnamefont {Ames}}, \bibinfo {author}
  {\bibfnamefont {C.}~\bibnamefont {Br\"{u}ne}}, \bibinfo {author}
  {\bibfnamefont {G.}~\bibnamefont {Tkachov}}, \bibinfo {author} {\bibfnamefont
  {H.}~\bibnamefont {Buhmann}}, \ and\ \bibinfo {author} {\bibfnamefont
  {L.~W.}\ \bibnamefont {Molenkamp}},\ }\href {\doibase
  10.1103/PhysRevX.3.021007} {\bibfield  {journal} {\bibinfo  {journal} {Phys.
  Rev. X}\ }\textbf {\bibinfo {volume} {3}},\ \bibinfo {pages} {021007}
  (\bibinfo {year} {2013})}\BibitemShut {NoStop}%
\bibitem [{\citenamefont {Hart}\ \emph {et~al.}()\citenamefont {Hart},
  \citenamefont {Ren}, \citenamefont {Wagner}, \citenamefont {Leubner},
  \citenamefont {M\"{u}hlbauer}, \citenamefont {Br\"{u}ne}, \citenamefont
  {Buhmann}, \citenamefont {Molenkamp},\ and\ \citenamefont
  {Yacoby}}]{21_Hart2014}%
  \BibitemOpen
  \bibfield  {author} {\bibinfo {author} {\bibfnamefont {S.}~\bibnamefont
  {Hart}}, \bibinfo {author} {\bibfnamefont {H.}~\bibnamefont {Ren}}, \bibinfo
  {author} {\bibfnamefont {T.}~\bibnamefont {Wagner}}, \bibinfo {author}
  {\bibfnamefont {P.}~\bibnamefont {Leubner}}, \bibinfo {author} {\bibfnamefont
  {M.}~\bibnamefont {M\"{u}hlbauer}}, \bibinfo {author} {\bibfnamefont
  {C.}~\bibnamefont {Br\"{u}ne}}, \bibinfo {author} {\bibfnamefont
  {H.}~\bibnamefont {Buhmann}}, \bibinfo {author} {\bibfnamefont {L.~W.}\
  \bibnamefont {Molenkamp}}, \ and\ \bibinfo {author} {\bibfnamefont
  {A.}~\bibnamefont {Yacoby}},\ }\href {\doibase 10.1038/nphys3036} {\bibinfo
  {journal} {Nat. Phys. (2014)}\ ,\ \bibinfo {pages} {doi:
  10.1038/nphys3036}}\BibitemShut {NoStop}%
\bibitem [{\citenamefont {Dynes}\ and\ \citenamefont
  {Fulton}(1971)}]{22_Dynes1971}%
  \BibitemOpen
\bibfield  {journal} {  }\bibfield  {author} {\bibinfo {author} {\bibfnamefont
  {R.}~\bibnamefont {Dynes}}\ and\ \bibinfo {author} {\bibfnamefont
  {T.}~\bibnamefont {Fulton}},\ }\href {\doibase 10.1103/PhysRevB.3.3015}
  {\bibfield  {journal} {\bibinfo  {journal} {Phys. Rev. B}\ }\textbf {\bibinfo
  {volume} {3}},\ \bibinfo {pages} {3015} (\bibinfo {year} {1971})}\BibitemShut
  {NoStop}%
\bibitem [{\citenamefont {Nichele}\ \emph {et~al.}(2014)\citenamefont
  {Nichele}, \citenamefont {Pal}, \citenamefont {Pietsch}, \citenamefont {Ihn},
  \citenamefont {Ensslin}, \citenamefont {Charpentier},\ and\ \citenamefont
  {Wegscheider}}]{23_Nichele2014}%
  \BibitemOpen
  \bibfield  {author} {\bibinfo {author} {\bibfnamefont {F.}~\bibnamefont
  {Nichele}}, \bibinfo {author} {\bibfnamefont {A.~N.}\ \bibnamefont {Pal}},
  \bibinfo {author} {\bibfnamefont {P.}~\bibnamefont {Pietsch}}, \bibinfo
  {author} {\bibfnamefont {T.}~\bibnamefont {Ihn}}, \bibinfo {author}
  {\bibfnamefont {K.}~\bibnamefont {Ensslin}}, \bibinfo {author} {\bibfnamefont
  {C.}~\bibnamefont {Charpentier}}, \ and\ \bibinfo {author} {\bibfnamefont
  {W.}~\bibnamefont {Wegscheider}},\ }\href {\doibase
  10.1103/PhysRevLett.112.036802} {\bibfield  {journal} {\bibinfo  {journal}
  {Phys. Rev. Lett.}\ }\textbf {\bibinfo {volume} {112}},\ \bibinfo {pages}
  {036802} (\bibinfo {year} {2014})}\BibitemShut {NoStop}%
\bibitem [{\citenamefont {Thinkham}(1996)}]{24_Thinkham1996}%
  \BibitemOpen
  \bibfield  {author} {\bibinfo {author} {\bibfnamefont {M.}~\bibnamefont
  {Thinkham}},\ }\href@noop {} {\emph {\bibinfo {title} {{Introduction to
  Superconductivity}}}}\ (\bibinfo  {publisher} {McGraw-Hill},\ \bibinfo {year}
  {1996})\BibitemShut {NoStop}%
\bibitem [{\citenamefont {Wang}\ \emph {et~al.}()\citenamefont {Wang},
  \citenamefont {Liu}, \citenamefont {Zhang}, \citenamefont {Samarth},
  \citenamefont {Zhang},\ and\ \citenamefont {Liu}}]{25_Wang2013}%
  \BibitemOpen
  \bibfield  {author} {\bibinfo {author} {\bibfnamefont {Q.}~\bibnamefont
  {Wang}}, \bibinfo {author} {\bibfnamefont {X.}~\bibnamefont {Liu}}, \bibinfo
  {author} {\bibfnamefont {H.-J.}\ \bibnamefont {Zhang}}, \bibinfo {author}
  {\bibfnamefont {N.}~\bibnamefont {Samarth}}, \bibinfo {author} {\bibfnamefont
  {S.-C.}\ \bibnamefont {Zhang}}, \ and\ \bibinfo {author} {\bibfnamefont
  {C.-X.}\ \bibnamefont {Liu}},\ }\href {http://arxiv.org/abs/1311.4113} {\
  }\Eprint {http://arxiv.org/abs/1311.4113} {arXiv:1311.4113} \BibitemShut
  {NoStop}%
\bibitem [{\citenamefont {Fu}\ and\ \citenamefont {Kane}(2009)}]{26_Fu2009}%
  \BibitemOpen
  \bibfield  {author} {\bibinfo {author} {\bibfnamefont {L.}~\bibnamefont
  {Fu}}\ and\ \bibinfo {author} {\bibfnamefont {C.}~\bibnamefont {Kane}},\
  }\href {\doibase 10.1103/PhysRevB.79.161408} {\bibfield  {journal} {\bibinfo
  {journal} {Phys. Rev. B}\ }\textbf {\bibinfo {volume} {79}},\ \bibinfo
  {pages} {161408} (\bibinfo {year} {2009})}\BibitemShut {NoStop}%
\bibitem [{\citenamefont {Beenakker}\ \emph {et~al.}(2013)\citenamefont
  {Beenakker}, \citenamefont {Pikulin}, \citenamefont {Hyart}, \citenamefont
  {Schomerus},\ and\ \citenamefont {Dahlhaus}}]{27_Beenakker2013}%
  \BibitemOpen
  \bibfield  {author} {\bibinfo {author} {\bibfnamefont {C.~W.~J.}\
  \bibnamefont {Beenakker}}, \bibinfo {author} {\bibfnamefont {D.~I.}\
  \bibnamefont {Pikulin}}, \bibinfo {author} {\bibfnamefont {T.}~\bibnamefont
  {Hyart}}, \bibinfo {author} {\bibfnamefont {H.}~\bibnamefont {Schomerus}}, \
  and\ \bibinfo {author} {\bibfnamefont {J.~P.}\ \bibnamefont {Dahlhaus}},\
  }\href {\doibase 10.1103/PhysRevLett.110.017003} {\bibfield  {journal}
  {\bibinfo  {journal} {Phys. Rev. Lett.}\ }\textbf {\bibinfo {volume} {110}},\
  \bibinfo {pages} {017003} (\bibinfo {year} {2013})}\BibitemShut {NoStop}%
\bibitem [{\citenamefont {Lee}\ \emph {et~al.}()\citenamefont {Lee},
  \citenamefont {Michaeli}, \citenamefont {Alicea},\ and\ \citenamefont
  {Yacoby}}]{28_Lee2014}%
  \BibitemOpen
  \bibfield  {author} {\bibinfo {author} {\bibfnamefont {S.-P.}\ \bibnamefont
  {Lee}}, \bibinfo {author} {\bibfnamefont {K.}~\bibnamefont {Michaeli}},
  \bibinfo {author} {\bibfnamefont {J.}~\bibnamefont {Alicea}}, \ and\ \bibinfo
  {author} {\bibfnamefont {A.}~\bibnamefont {Yacoby}},\ }\href
  {http://arxiv.org/abs/1403.2747} {\ }\Eprint {http://arxiv.org/abs/1403.2747}
  {arXiv:1403.2747} \BibitemShut {NoStop}%
\bibitem [{\citenamefont {Nguyen}\ \emph {et~al.}(2013)\citenamefont {Nguyen},
  \citenamefont {Aref}, \citenamefont {Kauppila}, \citenamefont {Meschke},
  \citenamefont {Winkelmann}, \citenamefont {Courtois},\ and\ \citenamefont
  {Pekola}}]{29_Nguyen2013}%
  \BibitemOpen
  \bibfield  {author} {\bibinfo {author} {\bibfnamefont {H.~Q.}\ \bibnamefont
  {Nguyen}}, \bibinfo {author} {\bibfnamefont {T.}~\bibnamefont {Aref}},
  \bibinfo {author} {\bibfnamefont {V.~J.}\ \bibnamefont {Kauppila}}, \bibinfo
  {author} {\bibfnamefont {M.}~\bibnamefont {Meschke}}, \bibinfo {author}
  {\bibfnamefont {C.~B.}\ \bibnamefont {Winkelmann}}, \bibinfo {author}
  {\bibfnamefont {H.}~\bibnamefont {Courtois}}, \ and\ \bibinfo {author}
  {\bibfnamefont {J.~P.}\ \bibnamefont {Pekola}},\ }\href {\doibase
  10.1088/1367-2630/15/8/085013} {\bibfield  {journal} {\bibinfo  {journal}
  {New J. Phys.}\ }\textbf {\bibinfo {volume} {15}},\ \bibinfo {pages} {085013}
  (\bibinfo {year} {2013})}\BibitemShut {NoStop}%
\end{thebibliography}%


\begin{thebibliography}{11}%
\normalsize
\makeatletter
\providecommand \@ifxundefined [1]{%
 \@ifx{#1\undefined}
}%
\providecommand \@ifnum [1]{%
 \ifnum #1\expandafter \@firstoftwo
 \else \expandafter \@secondoftwo
 \fi
}%
\providecommand \@ifx [1]{%
 \ifx #1\expandafter \@firstoftwo
 \else \expandafter \@secondoftwo
 \fi
}%
\providecommand \natexlab [1]{#1}%
\providecommand \enquote  [1]{``#1''}%
\providecommand \bibnamefont  [1]{#1}%
\providecommand \bibfnamefont [1]{#1}%
\providecommand \citenamefont [1]{#1}%
\providecommand \href@noop [0]{\@secondoftwo}%
\providecommand \href [0]{\begingroup \@sanitize@url \@href}%
\providecommand \@href[1]{\@@startlink{#1}\@@href}%
\providecommand \@@href[1]{\endgroup#1\@@endlink}%
\providecommand \@sanitize@url [0]{\catcode `\\12\catcode `\$12\catcode
  `\&12\catcode `\#12\catcode `\^12\catcode `\_12\catcode `\%12\relax}%
\providecommand \@@startlink[1]{}%
\providecommand \@@endlink[0]{}%
\providecommand \url  [0]{\begingroup\@sanitize@url \@url }%
\providecommand \@url [1]{\endgroup\@href {#1}{\urlprefix }}%
\providecommand \urlprefix  [0]{URL }%
\providecommand \Eprint [0]{\href }%
\providecommand \doibase [0]{http://dx.doi.org/}%
\providecommand \selectlanguage [0]{\@gobble}%
\providecommand \bibinfo  [0]{\@secondoftwo}%
\providecommand \bibfield  [0]{\@secondoftwo}%
\providecommand \translation [1]{[#1]}%
\providecommand \BibitemOpen [0]{}%
\providecommand \bibitemStop [0]{}%
\providecommand \bibitemNoStop [0]{.\EOS\space}%
\providecommand \EOS [0]{\spacefactor3000\relax}%
\providecommand \BibitemShut  [1]{\csname bibitem#1\endcsname}%
\let\auto@bib@innerbib\@empty
\bibitem [{\citenamefont {Dynes}\ and\ \citenamefont
  {Fulton}(1971)}]{01_Dynes1971}%
  \BibitemOpen
  \bibfield  {author} {\bibinfo {author} {\bibfnamefont {R.}~\bibnamefont
  {Dynes}}\ and\ \bibinfo {author} {\bibfnamefont {T.}~\bibnamefont {Fulton}},\
  }\href {\doibase 10.1103/PhysRevB.3.3015} {\bibfield  {journal} {\bibinfo
  {journal} {Phys. Rev. B}\ }\textbf {\bibinfo {volume} {3}},\ \bibinfo {pages}
  {3015} (\bibinfo {year} {1971})}\BibitemShut {NoStop}%
\bibitem [{\citenamefont {Hart}\ \emph {et~al.}()\citenamefont {Hart},
  \citenamefont {Ren}, \citenamefont {Wagner}, \citenamefont {Leubner},
  \citenamefont {M\"{u}hlbauer}, \citenamefont {Br\"{u}ne}, \citenamefont
  {Buhmann}, \citenamefont {Molenkamp},\ and\ \citenamefont
  {Yacoby}}]{02_Hart2014}%
  \BibitemOpen
  \bibfield  {author} {\bibinfo {author} {\bibfnamefont {S.}~\bibnamefont
  {Hart}}, \bibinfo {author} {\bibfnamefont {H.}~\bibnamefont {Ren}}, \bibinfo
  {author} {\bibfnamefont {T.}~\bibnamefont {Wagner}}, \bibinfo {author}
  {\bibfnamefont {P.}~\bibnamefont {Leubner}}, \bibinfo {author} {\bibfnamefont
  {M.}~\bibnamefont {M\"{u}hlbauer}}, \bibinfo {author} {\bibfnamefont
  {C.}~\bibnamefont {Br\"{u}ne}}, \bibinfo {author} {\bibfnamefont
  {H.}~\bibnamefont {Buhmann}}, \bibinfo {author} {\bibfnamefont {L.~W.}\
  \bibnamefont {Molenkamp}}, \ and\ \bibinfo {author} {\bibfnamefont
  {A.}~\bibnamefont {Yacoby}},\ }\href {\doibase 10.1038/nphys3036} {\bibinfo
  {journal} {Nat. Phys. (2014)}\ ,\ \bibinfo {pages} {doi:
  10.1038/nphys3036}}\BibitemShut {NoStop}%
\bibitem [{\citenamefont {Jiang}\ \emph {et~al.}(2011)\citenamefont {Jiang},
  \citenamefont {Pekker}, \citenamefont {Alicea}, \citenamefont {Refael},
  \citenamefont {Oreg},\ and\ \citenamefont {von Oppen}}]{03_Jiang2011}%
  \BibitemOpen
\bibfield  {journal} {  }\bibfield  {author} {\bibinfo {author} {\bibfnamefont
  {L.}~\bibnamefont {Jiang}}, \bibinfo {author} {\bibfnamefont
  {D.}~\bibnamefont {Pekker}}, \bibinfo {author} {\bibfnamefont
  {J.}~\bibnamefont {Alicea}}, \bibinfo {author} {\bibfnamefont
  {G.}~\bibnamefont {Refael}}, \bibinfo {author} {\bibfnamefont
  {Y.}~\bibnamefont {Oreg}}, \ and\ \bibinfo {author} {\bibfnamefont
  {F.}~\bibnamefont {von Oppen}},\ }\href {\doibase
  10.1103/PhysRevLett.107.236401} {\bibfield  {journal} {\bibinfo  {journal}
  {Phys. Rev. Lett.}\ }\textbf {\bibinfo {volume} {107}},\ \bibinfo {pages}
  {236401} (\bibinfo {year} {2011})}\BibitemShut {NoStop}%
\bibitem [{\citenamefont {Pikulin}\ and\ \citenamefont
  {Nazarov}(2012)}]{04_Pikulin2012}%
  \BibitemOpen
  \bibfield  {author} {\bibinfo {author} {\bibfnamefont {D.~I.}\ \bibnamefont
  {Pikulin}}\ and\ \bibinfo {author} {\bibfnamefont {Y.~V.}\ \bibnamefont
  {Nazarov}},\ }\href {\doibase 10.1103/PhysRevB.86.140504} {\bibfield
  {journal} {\bibinfo  {journal} {Phys. Rev. B}\ }\textbf {\bibinfo {volume}
  {86}},\ \bibinfo {pages} {140504} (\bibinfo {year} {2012})}\BibitemShut
  {NoStop}%
\bibitem [{\citenamefont {San-Jose}\ \emph {et~al.}(2012)\citenamefont
  {San-Jose}, \citenamefont {Prada},\ and\ \citenamefont
  {Aguado}}]{05_San-Jose2012}%
  \BibitemOpen
  \bibfield  {author} {\bibinfo {author} {\bibfnamefont {P.}~\bibnamefont
  {San-Jose}}, \bibinfo {author} {\bibfnamefont {E.}~\bibnamefont {Prada}}, \
  and\ \bibinfo {author} {\bibfnamefont {R.}~\bibnamefont {Aguado}},\ }\href
  {\doibase 10.1103/PhysRevLett.108.257001} {\bibfield  {journal} {\bibinfo
  {journal} {Phys. Rev. Lett.}\ }\textbf {\bibinfo {volume} {108}},\ \bibinfo
  {pages} {257001} (\bibinfo {year} {2012})}\BibitemShut {NoStop}%
\bibitem [{\citenamefont {Dom\'{\i}nguez}\ \emph {et~al.}(2012)\citenamefont
  {Dom\'{\i}nguez}, \citenamefont {Hassler},\ and\ \citenamefont
  {Platero}}]{06_Dominguez2012}%
  \BibitemOpen
  \bibfield  {author} {\bibinfo {author} {\bibfnamefont {F.}~\bibnamefont
  {Dom\'{\i}nguez}}, \bibinfo {author} {\bibfnamefont {F.}~\bibnamefont
  {Hassler}}, \ and\ \bibinfo {author} {\bibfnamefont {G.}~\bibnamefont
  {Platero}},\ }\href {\doibase 10.1103/PhysRevB.86.140503} {\bibfield
  {journal} {\bibinfo  {journal} {Phys. Rev. B}\ }\textbf {\bibinfo {volume}
  {86}},\ \bibinfo {pages} {140503} (\bibinfo {year} {2012})}\BibitemShut
  {NoStop}%
\bibitem [{\citenamefont {Du}\ \emph {et~al.}()\citenamefont {Du},
  \citenamefont {Knez}, \citenamefont {Sullivan},\ and\ \citenamefont
  {Du}}]{07_Du2013}%
  \BibitemOpen
  \bibfield  {author} {\bibinfo {author} {\bibfnamefont {L.}~\bibnamefont
  {Du}}, \bibinfo {author} {\bibfnamefont {I.}~\bibnamefont {Knez}}, \bibinfo
  {author} {\bibfnamefont {G.}~\bibnamefont {Sullivan}}, \ and\ \bibinfo
  {author} {\bibfnamefont {R.-R.}\ \bibnamefont {Du}},\ }\href
  {http://arxiv.org/abs/1306.1925} {\ }\Eprint {http://arxiv.org/abs/1306.1925}
  {arXiv:1306.1925} \BibitemShut {NoStop}%
\bibitem [{\citenamefont {Thinkham}(1996)}]{08_Thinkham1996}%
  \BibitemOpen
  \bibfield  {author} {\bibinfo {author} {\bibfnamefont {M.}~\bibnamefont
  {Thinkham}},\ }\href@noop {} {\emph {\bibinfo {title} {{Introduction to
  Superconductivity}}}}\ (\bibinfo  {publisher} {McGraw-Hill},\ \bibinfo {year}
  {1996})\BibitemShut {NoStop}%
\bibitem [{\citenamefont {Doh}\ \emph {et~al.}(2005)\citenamefont {Doh},
  \citenamefont {van Dam}, \citenamefont {Roest}, \citenamefont {Bakkers},
  \citenamefont {Kouwenhoven},\ and\ \citenamefont {{De
  Franceschi}}}]{09_Doh2005}%
  \BibitemOpen
  \bibfield  {author} {\bibinfo {author} {\bibfnamefont {Y.-J.}\ \bibnamefont
  {Doh}}, \bibinfo {author} {\bibfnamefont {J.~A.}\ \bibnamefont {van Dam}},
  \bibinfo {author} {\bibfnamefont {A.~L.}\ \bibnamefont {Roest}}, \bibinfo
  {author} {\bibfnamefont {E.~P. A.~M.}\ \bibnamefont {Bakkers}}, \bibinfo
  {author} {\bibfnamefont {L.~P.}\ \bibnamefont {Kouwenhoven}}, \ and\ \bibinfo
  {author} {\bibfnamefont {S.}~\bibnamefont {{De Franceschi}}},\ }\href
  {\doibase 10.1126/science.1113523} {\bibfield  {journal} {\bibinfo  {journal}
  {Science}\ }\textbf {\bibinfo {volume} {309}},\ \bibinfo {pages} {272}
  (\bibinfo {year} {2005})}\BibitemShut {NoStop}%
\bibitem [{\citenamefont {Charpentier}\ \emph {et~al.}(2013)\citenamefont
  {Charpentier}, \citenamefont {Fält}, \citenamefont {Reichl}, \citenamefont
  {Nichele}, \citenamefont {{Nath Pal}}, \citenamefont {Pietsch}, \citenamefont
  {Ihn}, \citenamefont {Ensslin},\ and\ \citenamefont
  {Wegscheider}}]{10_Charpentier2013}%
  \BibitemOpen
  \bibfield  {author} {\bibinfo {author} {\bibfnamefont {C.}~\bibnamefont
  {Charpentier}}, \bibinfo {author} {\bibfnamefont {S.}~\bibnamefont {Fält}},
  \bibinfo {author} {\bibfnamefont {C.}~\bibnamefont {Reichl}}, \bibinfo
  {author} {\bibfnamefont {F.}~\bibnamefont {Nichele}}, \bibinfo {author}
  {\bibfnamefont {A.}~\bibnamefont {{Nath Pal}}}, \bibinfo {author}
  {\bibfnamefont {P.}~\bibnamefont {Pietsch}}, \bibinfo {author} {\bibfnamefont
  {T.}~\bibnamefont {Ihn}}, \bibinfo {author} {\bibfnamefont {K.}~\bibnamefont
  {Ensslin}}, \ and\ \bibinfo {author} {\bibfnamefont {W.}~\bibnamefont
  {Wegscheider}},\ }\href {\doibase 10.1063/1.4821037} {\bibfield  {journal}
  {\bibinfo  {journal} {Appl. Phys. Lett.}\ }\textbf {\bibinfo {volume}
  {103}},\ \bibinfo {pages} {112102} (\bibinfo {year} {2013})}\BibitemShut
  {NoStop}%
\bibitem [{\citenamefont {Yu}\ \emph {et~al.}(2005)\citenamefont {Yu},
  \citenamefont {Singh}, \citenamefont {Liu}, \citenamefont {Wu}, \citenamefont
  {Hu}, \citenamefont {Durand}, \citenamefont {Bulman}, \citenamefont
  {Rowell},\ and\ \citenamefont {Newman}}]{11_Yu2005}%
  \BibitemOpen
  \bibfield  {author} {\bibinfo {author} {\bibfnamefont {L.}~\bibnamefont
  {Yu}}, \bibinfo {author} {\bibfnamefont {R.}~\bibnamefont {Singh}}, \bibinfo
  {author} {\bibfnamefont {H.}~\bibnamefont {Liu}}, \bibinfo {author}
  {\bibfnamefont {S.}~\bibnamefont {Wu}}, \bibinfo {author} {\bibfnamefont
  {R.}~\bibnamefont {Hu}}, \bibinfo {author} {\bibfnamefont {D.}~\bibnamefont
  {Durand}}, \bibinfo {author} {\bibfnamefont {J.}~\bibnamefont {Bulman}},
  \bibinfo {author} {\bibfnamefont {J.}~\bibnamefont {Rowell}}, \ and\ \bibinfo
  {author} {\bibfnamefont {N.}~\bibnamefont {Newman}},\ }\href {\doibase
  10.1109/TASC.2005.844126} {\bibfield  {journal} {\bibinfo  {journal} {IEEE
  Trans. Appl. Supercond.}\ }\textbf {\bibinfo {volume} {15}},\ \bibinfo
  {pages} {44} (\bibinfo {year} {2005})}\BibitemShut {NoStop}%
\end{thebibliography}
